\documentclass[a4paper, 11pt]{article}
\usepackage[left=2.cm,right=2.cm, top=2.5cm,bottom=2.5cm]{geometry}
\usepackage{amsmath}
\usepackage{amsfonts}
\usepackage{amssymb}
\usepackage{amsthm} 
\usepackage{multicol}
\usepackage{graphicx}
\usepackage{enumerate}
\usepackage[inline,shortlabels]{enumitem}
\usepackage{bbm}
\usepackage[usenames,dvipsnames]{color}
\usepackage[colorlinks=true,linkcolor=blue, citecolor=blue, urlcolor=blue]{hyperref}
\usepackage{fancyhdr}
\usepackage{epsfig}
\usepackage{epstopdf}
\usepackage{caption}
\usepackage[textfont=footnotesize,justification=centering]{subcaption}
\usepackage{multirow}
\usepackage{appendix}
\usepackage{float}
\usepackage{dsfont}
\usepackage{csquotes} 
\usepackage{comment}
\usepackage[multiple]{footmisc}
\usepackage{xcolor}

\usepackage{fancyhdr}
\fancyheadoffset{0cm}

\fancypagestyle{plain}{
\fancyhf{}
\fancyheadoffset{0cm}

\fancyfoot[C]{\thepage}
}

\newtheorem*{theorem-non}{Theorem}
\newtheorem{theorem}{Theorem}

\newtheorem{proposition}{Proposition}

\newtheorem{corollary}{Corollary}
\newtheorem{remark}{Remark}

\usepackage[sort]{natbib}
\bibpunct{(}{)}{;}{a}{,}{,}

\renewcommand{\footnoterule}{%
  \kern -3pt
  \hrule width 1\textwidth height 0.1pt
  \kern 2pt
}

\begin{document}
\title{Reference-dependent asset pricing \\ with a stochastic consumption-dividend ratio\\[.5cm]
}

\author{Luca De Gennaro Aquino\thanks{Department of Mathematical Sciences, University of Copenhagen, Copenhagen, Denmark.} \and Xuedong He\thanks{Department of Systems Engineering and Engineering Management, The Chinese University of Hong Kong, Hong Kong.} \and Moris S. Strub\thanks{Warwick Business School, University of Warwick, Coventry, United Kingdom.} \and Yuting Yang\thanks{Department of Systems Engineering and Engineering Management, The Chinese University of Hong Kong, Hong Kong.} }

\maketitle

\begin{abstract}
We study a discrete-time consumption-based capital asset pricing model under expectations-based reference-dependent
preferences. More precisely, we consider an endowment economy populated by a representative agent who derives utility from current consumption and from gains and losses in consumption with respect to a forward-looking, stochastic reference point.  First, we consider a general model in which the agent's preferences include both contemporaneous gain-loss utility, that is, utility from the difference between current consumption and previously held expectations about current consumption, and prospective gain-loss utility, that is, utility from the difference between intertemporal beliefs about future consumption. A semi-closed form solution for equilibrium asset prices is derived for this case.
We then specialize to a model in which the agent derives contemporaneous gain-loss utility only, obtaining equilibrium asset prices in closed form.
Extensive numerical experiments show that, with plausible values of risk aversion and loss aversion, our models can generate equity premia that match empirical estimates. Interestingly, the models turn out to be consistent with some well-known empirical facts, namely procyclical variation in the price-dividend ratio and countercyclical variation in the conditional expected equity premium and in the conditional volatility of the equity premium. Furthermore, we find that prospective gain-loss utility is necessary for the model to predict reasonable values of the price-dividend ratio.
\end{abstract}

{\textbf{\\Keywords: }consumption-based asset pricing, gain-loss utility, reference-dependent preferences,\\ consumption-dividend ratio}

\section{Introduction}

Pioneered by \cite{Lucas1978:Econometrica}, the classical consumption-based capital asset pricing model (hereafter, CCAPM) states that, in equilibrium, the expected return of an asset is equal to the risk-free rate plus a premium for consumption risk that depends on the covariance between the asset returns and the stochastic discount factor - or, equivalently, the intertemporal marginal rate of substitution. 
Assets whose returns are low when the marginal utility of consumption is high are perceived as more risky and investors thus expect to receive a larger equity risk premium to hold them.\footnote{Prior to the work of Lucas, \cite{Rubinstein1976:BellJournalEconomics} had already formalized the idea that asset prices can be represented by a stochastic discount factor in the form of an intertemporal marginal rate of substitution.}\textsuperscript{,}\footnote{The continuous-time version of the CCAPM is due to \cite{Breeden1979:JFE}.}

The financial economics community attentively scrutinized the implications of the CCAPM, and evidence of its inability to match financial market data started to emerge. 
In particular, \cite{Mehra1985:JME} asserted that a frictionless, pure exchange economy with a representative expected utility agent implies values of the equity risk premium that are too low when calibrated to historical consumption data for plausible values of the discount factor and coefficient of relative risk aversion. 
This observation came to be known as the \textit{equity premium puzzle}. 
In relation to this argument, \cite{Weil1989:JME} observed that even if one used an implausibly high risk aversion degree to fit the empirical equity premium, the CCAPM would yield an unrealistically high risk-free rate. 
This issue was labeled as the \textit{risk-free rate
puzzle}.\footnote{\cite{HansenJagannathan1991:JPE} provided an explanation of why the CCAPM could not address the equity premium and risk-free rate puzzles by calculating bounds on the mean and standard deviation of intertemporal marginal rates of substitution. \cite{CochraneHansen1992:NBER}, \cite{Burnside1994:JBES}, and \cite{CecchettiLamMark1994:JF} further developed these ideas, together with a series of statistical methods for testing various implications of asset pricing models.}\textsuperscript{,}\footnote{Other empirical findings that could not be explained by the standard CCAPM - among others, the high volatility and predictability at long horizons of stock returns - were identified by \cite{Shiller1980:AER}, \cite{LeroyPorter1981}, \cite{FamaFrench1989:JFE}, and \cite{Campbell1991:EJ, Campbell1999:HM}.}

Over the last 50 years, a plethora of extensions and variations of Lucas' CCAPM have been proposed, in part to attempt to solve the equity premium puzzle, possibly in conjunction with other stock market anomalies or stylized facts. We summarize here the main trends, referring the reader also to the diligent surveys in \cite{Kocherlakota1996:JEL}, \cite{Campbell2003:HandbookEconomicsFinance}, \cite{DonaldsonMehra2008:HandbookERP} and \cite{MehraPrescott2008:HandbookERP}.

Fundamentally, researchers have dealt with the puzzles in question in one of the following ways, or combination of these: (i) by introducing some form of trading frictions, constraints or market incompleteness; (ii) by allowing for rare, disastrous scenarios that produce a sharp drop in consumption (such as market crashes or recessions); or (iii) by modifying the preferences of the representative agent. 
One can argue that, in all these cases, the goal is to make agents more averse to undesirable consumption outcomes and lessen the connection between asset returns and marginal rates of intertemporal substitution.

For example, within the first stream, \cite{ScheinkmanWeiss1986:Econometrica} focused on borrowing constraints, \cite{Zeldes1989:JPE} on liquidity constraints, \cite{GrossmanLaroque1990:Econometrica}, \cite{HeModest1995:JPE}, and \cite{Luttmer1996:Econometrica} on transaction costs. Abstracting from the representative-consumer framework, \cite{Mankiw1986:JFE} considered an incomplete market where ex-ante identical agents are affected to different extents by consumption shocks, while \cite{ConstantinidesDuffie1996:JPE} proposed an heterogeneous-agent model with heteroscedastic labor income shocks. 
Within the second stream, the seminal contribution is by \cite{Rietz1988:JME}, with a more recent revitalization by \cite{Barro2006:QJE}.

The third stream of the literature, to which the present paper belongs, covers a large spectrum of theories. 
\cite{Constantinides1990:JPE}, followed by \cite{FersonConstantinides1991:JFE}, \cite{Chapman1998:Econometrica}, \cite{CampbellCochrane1999:JPE}, employed models in which individuals form internal habits so that their utility depends not only on current consumption but also on their own past levels of consumption. \cite{Abel1990:AER} combined internal habit persistence with  an external reference point determined by historical aggregate average consumption, often referred as the \enquote{catching up with the Joneses} specification. 
\cite{Gul1991:Econometrica},  \cite{BonomoGarciaMeddahiTedongap2010:RFS}, \cite{CampanaleCastroClementi2010:RED}, \cite{RoutledgeZin2010:JF} developed the concept of disappointment aversion, according to which agents overweight lower-tail, disappointing outcomes compared to expected utility theory.  
Building on the foundations laid out by \cite{KrepsPorteus1978:Econometrica}, \cite{EpsteinZin1989:Econometrica, EpsteinZin1991:JPE} and \cite{Weil1989:JME, Weil1990:QJE} proposed and applied recursive utility preferences disentangling relative risk aversion and elasticity of intertemporal substitution. 
Among others, \cite{BenartziThaler1995:QJE}, \cite{BarberisHuangSantos2001:QJE}, and \cite{VanBilsenLaegenNijman2020:MS}, in addition to utility of consumption, consider gains and losses in financial wealth relative to a reference point, one of the key tenets of \cite{KahnemanTversky1979:Econometrica}'s prospect theory.\footnote{Expanding on this, \cite{BarberisJinWang2021:JF} recently stress tested prospect theory against 21 stock market anomalies. Cumulative prospect theory \citep{TverskyKahneman1992:JRU} has instead been examined by \cite{Gomes2005:JB}, \cite{BarberisXiong2009:JF}, \cite{DeGiorgiLegg2012:JEDC}, \cite{Andries2012}. 
}\textsuperscript{,}\footnote{Papers merging features from various theories include \cite{Yogo2008:JBES}, who mixed prospect theory and habit formation with a slowly time-varying habit, \cite{BarberisHuang2009:JEDC}, who incorporated narrow framing into recursive utility, and \cite{EasleyYang2015:JET} and \cite{GuoHe2017:JEDC}, who considered a model with both Epstein-Zin and cumulative prospect theory agents. We also mention the works of  \cite{Weitzman2007:AER} and \cite{BarillasHansenSargent2009:JET}, who departed from rational expectations and shed light on the impact of model uncertainty for equilibrium asset pricing.}

In the above-mentioned reference-dependent models, the agent's benchmark is supposed to be backward-looking, say as function of past wealth levels. \cite{KoszegiRabin2006:QJE, KoszegiRabin2007:AER} advanced the alternative idea of a forward-looking reference point defined by rational expectations about future outcomes. For consumption plans, this reference point would then be determined by the agent's expectation about her future consumption, given current information and beliefs; see  \cite{KoszegiRabin2009:AER}. This yields that, at each time $t$, the agent's utility $\tilde{U}_t$ can be written as the sum of a standard consumption utility and a series of $T+1$ terms depending on the updated beliefs about contemporaneous and future consumption:
\begin{eqnarray} \label{KoszegiRabin_GeneralPreferences}
\begin{aligned}
\tilde{U}_t& = m(C_t) +\sum_{s=0}^{T}\beta^s  \mathbb{E}_t \left[\mu \left(m(\tilde C_{t,t+s})-m(\tilde{C}_{t-1,t+s})\right)\right]. 
\end{aligned}
\end{eqnarray}
We will formalize all the ingredients in the sections below, so for now it should suffice to say that $m(\cdot)$ represents the consumption utility and $\mu(\cdot)$ the gain-loss utility, with $\tilde C_{\tau_{1},\tau_{2}}$ being the agent's belief held at time $\tau_{1}$ about her consumption at time $\tau_{2}$, $\tau_{1} \leq \tau_{2}$. 
Expected gain-loss utility of future consumption is discounted by a constant factor $\beta$. 

This class of preferences has been adopted for a consumption-based asset pricing model in \cite{Pagel2016:JEEA}, who studied an endowment economy à la Lucas (consumption and dividends are taken to be identical) with one representative Kőszegi-Rabin agent. 
Her analysis shows that loss averse agents tend to reduce the odds of unexpected cuts in consumption by postponing it to periods when their beliefs are less optimistic; see also \cite{Pagel2017:RES}. Therefore, to compensate for the uncertainty attached to painful fluctuations in consumption, agents must precautionarily save or invest larger shares of their wealth and demand higher equity premia. This aversion towards negative new information also appears in \cite{Pagel18:ECMA}, where the same preferences are applied in the context of a portfolio selection problem in which the agent can decide to monitor and rebalance his portfolio less frequently - or even pay for delegated portfolio management - as a way to reduce news-disutility. 
Further insights on the portfolio choice with Kőszegi-Rabin preferences have been derived by \cite{MengWeng18:MS,GuasoniMeirelesRodrigues2020:MOR, GuasoniMeirelesRodrigues2023:MF}.

The objective of the present paper is to extend the asset pricing model by \cite{Pagel2016:JEEA} in two aspects. 
First, in accordance with empirical evidence, we let consumption and dividends be driven by separate processes.\footnote{Annual data spanning the period 1929-2022, sourced from the U.S. Department of Commerce, Bureau of Economic Analysis, indicate that, on average, aggregate consumption exceeds dividends by a factor of 21 and is stochastic, with a range for this factor extending from approximately 10 to 31.} 
Second, we allow for the optimal consumption propensity (i.e., fraction of wealth consumed) to be wealth-dependent, as opposed to independent of wealth. Along the way, we provide some technical results on the existence of equilibrium solutions and comparative statics with respect to the preference parameters and state variables of the model. Further, in addition to the equity risk premium and the risk free rate, we obtain formulas for the stock price-dividend ratio that seem to be new.

We start our analysis by considering a general model (Model I) in which the agent derives both contemporaneous and prospective gain-loss utility  - setting $T \equiv \infty$ in the preference specification \eqref{KoszegiRabin_GeneralPreferences}. In this case, we calculate a set of first-order conditions that are satisfied by asset prices and derive a semi-closed form equilibrium solution for the risk-free rate, stock price-dividend ratio, and (excess) stock return. Also, under mild assumptions, we prove the existence of such a solution. 

Our numerical experiments exhibit that, with reasonable preference parameters, the general model is able to generate equilibrium prices that are close to the empirical estimates.  
Moreover, we find that the model is consistent with some well-known empirical facts, namely the procyclical variation in the price-dividend ratio (\cite{FamaFrench1989:JFE}) and the countercyclical variation in the conditional expected equity premium (\cite{CampbellShiller1988:JF, CampbellShiller1988:RFS}, \cite{FamaFrench1989:JFE}) and in the conditional volatility of the equity premium (\cite{BollerslevChouKroner1992:JE}). 
Alas, it appears difficult to match the historically low volatility and procyclical variation of risk-free rates (\cite{Fama1990:JME}). 

We then move on to focus on the special case (Model II) in which the agent derives contemporaneous gain-loss utility only - by setting $T \equiv 0$ in \eqref{KoszegiRabin_GeneralPreferences}. This case allows us to derive equilibrium asset prices in closed form. 

The tractability of Model II enables us to build additional intuition about the role of gain-loss utility for asset pricing. In particular, we show that whether a higher degree of loss aversion, or of the weight attributed to the contemporaneous gain-loss utility, leads to higher or lower equilibrium asset prices depends on the growth rate of consumption in the current period. Intuitively, if the growth rate of consumption is high, current consumption is more likely to be in the gain region, so the increment in the marginal utility of consumption in the current period - induced by a higher loss aversion degree - is smaller than the increment in the marginal utility of consumption in the next period. As a consequence, the agent defers consumption and saves or invests more now, thereby driving down (respectively, up) the equilibrium risk-free rate (price-dividend ratio). 

While some of the key model predictions of the general Model I hold for Model II as well, our numerics show that the special case without prospective gain-loss utility does not capture low risk-free rates and high price-dividend ratios. 

The rest of the paper is organized as follows. In  Section \ref{sec:GeneralModel}, we consider a general model in which the agent derives both contemporaneous and prospective gain-loss utility from consumption. In Section \ref{sec:BaselineModel}, we move to a special case in which the agent only derives contemporaneous gain-loss utility.  Numerics are conducted in Section \ref{sec:Numerics}. Section \ref{sec:Conclusions} concludes. Proofs and additional results are provided in the appendix.

\section{Model I: contemporaneous and prospective gain-loss utility}
\label{sec:GeneralModel}

\subsection{The market} \label{subsec:TheMarket}
We consider a discrete-time, pure exchange economy with one representative agent. 
The agent can trade two assets: a risk-free asset in zero net supply and a risky asset - a stock -  in unit net supply. In the time period between $t$ and $t+1$, the (gross) return rates of the risk-free asset and the stock are $R_{f,t+1}$ and $R_{S,t+1}$, respectively, while $D_{t}$ represents the dividends paid out by the stock at time $t$. Denoting by $S_{t}$ the stock price, the stock return is then 
\begin{align}
R_{S,t+1} = \frac{S_{t+1}+D_{t+1}}{S_t}.\label{eq:StockGrossReturn}
\end{align}
In contrast with standard Lucas-type models, our framework separates dividend and (aggregate) consumption streams, the latter of which we denote by $\bar{C}_{t}$. 
While different ways have been proposed in the literature to characterize this separation (see, e.g., \cite{CecchettiLamMark1993:JME}, \cite{BarberisHuangSantos2001:QJE}, \cite{BansalYaron2004:JF}), the majority of them specify a bivariate process for consumption growth and dividend growth rates. For reasons that will be more evident in Subsection \ref{subsec:GM_DefinitionEquilibrium}, we instead directly model the consumption growth rate and the consumption-dividend ratio.\footnote{The forecasting power of consumption growth and consumption-dividend ratio on excess stock returns has been documented by \cite{LettauLudvigson2001:JF, LettauLudvigson2005:JFE}, while   \cite{BekaertEngstromGrenadier2010:JEF} utilize these two state variables for predicting bond and stock returns in a model with an exogenous, stochastic habit level.}  Specifically, we assume that the consumption growth rate $\varepsilon_{c,t+1} := \dfrac{\bar{C}_{t+1}}{\bar{C}_{t}} $ follows a log-normal i.i.d. process,
\begin{equation}\label{eq:ConsGrowthLogNormal}
\log(\varepsilon_{c,t+1}) = \mu_c +  \sigma_c z_{t+1}, \quad z_{t} \sim \mathcal{N}(0,1) \mbox{ i.i.d.},
\end{equation}
for constant parameters $\mu_{c}, \sigma_{c}$, and that the consumption-dividend ratio $ Y_{t+1} := \dfrac{\bar{C}_{t+1}}{D_{t+1}} $ follows a log-normal AR(1) process,
\begin{equation}\label{eq:lognormalAR(1)}
\log(Y_{t+1}) = (1-\varphi)\kappa + \varphi\log(Y_{t}) +  \sigma_{y}\epsilon_{t+1}, \quad \epsilon_{t} \sim \mathcal{N}(0,1) \mbox{ i.i.d.}, 
\end{equation}
for constant parameters $\varphi, \kappa,\sigma_{y}$.  We refer to \cite{LettauWachter2007:JF} for a similar construction.

Two observations are in order. 
First, we emphasize that our theoretical results hold under more general distributional assumptions. In principle, besides a specific growth condition that will be mentioned later on, we only require the consumption growth rate to be i.i.d. over time and the consumption-dividend ratio to be a Markov chain. Second, we point out the following relation between the dividend growth rate $\varepsilon_{d,t+1} := \dfrac{D_{t+1}}{D_{t}}$, the consumption growth rate and the consumption-dividend ratio:
\begin{align}
\varepsilon_{d,t+1} =\varepsilon_{c,t+1} \dfrac{Y_{t}}{Y_{t+1}}.\label{eq:DivGrowthFromConDivRatio}
\end{align}
This relation will come in handy when we need to derive the properties of the dividend growth rate. Further details on these aspects and parameter calibration of the state variables are postponed to \mbox{Section \ref{sec:Numerics}}.

\subsection{Preferences}\label{subsec:GM_Preferences}

At each time $t$, the agent chooses the consumption $C_t$ and the percentage $\alpha_t$ of her post-consumption wealth to be invested in the stock. We denote by $C_t^{\star}$ the optimal consumption and $\alpha_t^{\star}$ the optimal allocation in the stock.

Upon receiving a labor (non-financial) income $L_{t} := \bar{C}_{t} - D_{t}$,  the agent's wealth $W_{t}$ thus evolves as
\begin{align}
W_{t+1} = (W_t-C_t)(\alpha_t R_{S,t+1}+(1-\alpha_t)R_{f,t+1})+L_{t+1}, \quad t = 0,1,\dots\label{eq:WealthEquation}
\end{align}

The agent's utility for a consumption stream $\{C_s\}_{s\geq t}$ is measured by a variant of the reference-dependent model proposed by \citet{KoszegiRabin2007:AER}. In this model, the agent evaluates  her consumption $C_s$  at time $s$ with respect to a reference point formed at time $s-1$. This reference point, denoted by $\tilde C_{s-1,s}$, is the agent's expectation held at time $s-1$ of her consumption at time $s$. Following \citet{KoszegiRabin2007:AER}, we assume that $\{\tilde C_{s-1,s}\}_{s\in \mathbb{N}}$ is independent of ${\cal F}_\infty$, where $\{\mathcal{F}_t\}_{t\in \mathbb{N}}$ denotes the filtration generated by the aggregate endowment and dividend processes, i.e., ${\cal F}_t=\sigma\{(\varepsilon_{c,s},\varepsilon_{d,s}): s\le t\}$. 
In particular, $\tilde C_{s-1,s}$ is independent of $C_s$, so the agent compares each outcome of $C_s$ with every possible outcome of the reference point.\footnote{An alternative preference specification where a prospect is evaluated against a reference point formed by endogenous expectations in the same state was proposed by \cite{DeGiorgiPost11:MS} and further studied in the context of portfolio optimization in complete markets in \cite{HeStrub2022:OR}.}
We denote by $\mathbb{E}_s [X] = \mathbb{E} [X \vert \mathcal{F}_s]$ the  conditional expectation of a prospect $X$ given $\mathcal{F}_s$.

At time $t$, the agent's cumulative utility is then given by
\begin{equation*}
\tilde V_t\left(\{C_r\}_{r\ge t}  \mid \{\tilde{C}_{r-1,r}\}_{r\ge t},\{\hat{C}_{r,r+i}\}_{r\ge t,i\ge 1},\{\hat{C}_{r-1,r+i}\}_{r\ge t,i\ge 1}\right) = \mathbb{E}_t\left[\sum_{s=t}^{\infty} \beta^{s-t} \tilde U_s\right],
\end{equation*}
where
\begin{eqnarray}\label{GM_preference}
\begin{aligned}
\tilde{U}_t&= U_t(C_t) + \gamma \sum_{s=1}^{\infty}\beta^s  \mathbb{E}_t \left[\mu \left(m(\hat C_{t,t+s})-m(\hat{C}_{t-1,t+s})\right)\right]\\
&= m(C_t)+b \mathbb{E}_t \left[ \mu \left( m( C_t) - m(\tilde {C}_{t-1,t}) \right) \right] + \gamma \sum_{s=1}^{\infty}\beta^s  \mathbb{E}_t \left[\mu \left(m(\hat C_{t,t+s})-m(\hat{C}_{t-1,t+s})\right)\right], \quad t \geq 0.\\[0.2cm]
\end{aligned}
\end{eqnarray}
We describe each term in \eqref{GM_preference} separately. The first term, $m(C_t)$, corresponds to the classical utility of consumption, with 
\begin{equation} \label{eq:power_utility_function} 
m(x) :=
\begin{cases}
\dfrac{x^{1-\theta}}{1-\theta}, & 0 < \theta \neq 1, \\
\log(x), & \theta = 1,
\end{cases}
\end{equation}
being a power utility function with relative risk aversion degree $\theta$.

The second term, $b\mathbb{E}_t \left[ \mu \left( m(C_t) - m(\tilde{C}_{t-1,t}) \right) \right]$, is called \textit{contemporaneous gain-loss utility} and measures the agent's utility of gains and losses in consumption. More precisely, at time $t-1$ the agent expects her future consumption at time $t$ to be $\tilde{C}_{t-1,t}$, while at time $t$ she actually consumes $C_t$. In terms of consumption utility, the gain or loss is therefore $ m(C_t) - m(\tilde{C}_{t-1,t})$. In order to model the intuition that the disutility of a loss is larger than the utility of a gain of the same size, we use the piecewise linear function
\begin{eqnarray}
\mu(x) :=
\begin{cases}
x, & x \geq 0,\\
\lambda x, & x < 0,
\end{cases}\label{eq:gainlossutilityfun}
\end{eqnarray}
for a constant $\lambda\ge 1$. We will refer to $\mu(\cdot)$ as the gain-loss utility function and $\lambda$ as the \textit{degree of loss aversion} (DLA). Finally, the parameter $b\ge 0$ in front of the expectation measures the relative weight of the contemporaneous gain-loss utility component versus the consumption utility component.

The last term, $\gamma \sum_{s=1}^{\infty}\beta^s  \mathbb{E}_t \left[\mu \left(m(\hat C_{t,t+s})-m(\hat{C}_{t-1,t+s})\right)\right]$, is called \textit{prospective gain-loss utility} and measures the utility from the difference between intertemporal beliefs about future consumption. Namely, for each $s\ge 1$, $\hat{C}_{t-1,t+s}$ and $\hat{C}_{t,t+s}$ represent the agent's expectation at time $t-1$ and at time $t$, respectively, of future consumption at time $t+s$. The evaluation of future gains or losses is then based on the quantity $\mu\left(m(\hat C_{t,t+s})-m(\hat{C}_{t-1,t+s})\right)$ and follows the same behavioral principles described in the previous paragraph. Here the constant $\gamma \geq 0$ measures the relative weight given to the prospective gain-loss utility component.\footnote{Clearly, when $b= \gamma = 0$, $\tilde{U}_t$ boils down to the classical consumption utility.}

We now discuss how the agent's expectation about her current consumption (i.e.,  $\tilde{C}_{t-1,t}$) and her future consumptions (i.e., $\hat{C}_{t-1,t+s}$ and $\hat C_{t,t+s}$) are created. For current consumption, we assume that at time $t-1$ the agent has rational expectations about her consumption at time $t$, which are then used as the reference point for her true consumption at time $t$. As a result, conditional on the information available at time $t-1$, the reference point $\tilde{C}_{t-1,t}$ follows the same distribution of $C_t^{\star}$, the amount that the agent plans to optimally consume at time $t$. 

For future consumptions, we need a preliminary observation. From time $t-1$ to $t$, the agent's expectations are updated in two respects: First, through the arrival of new information by means of the consumption growth rate $\varepsilon_{c,t}$ and the dividend growth rate $\varepsilon_{d,t}$. Second, the agent's decisions at the current time $t$, namely the consumption $C_t$ and investment in the stock $\alpha_t$, affect the wealth at time $t+1$ and, in consequence, the agent's expectation about future consumption: the more wealth the agent expects to have at time $t+1$, the more she expects to consume in the future. A (fully) rational expectation assumption should then be based on a general consumption rule that depends on the wealth level and other state variables (we will see this in more detail in Subsection \ref{subsec:GM_DefinitionEquilibrium}). For tractability, we instead assume that the agent is of bounded rationality with respect to future consumption, so that she expects her future selves to follow a linear consumption rule. Moreover, we assume that the agent expects the growth rate of her wealth at future times to be independent of the wealth level. In other words,
\begin{align}
&\hat C_{t,t+i} = W_{t+i} \,\hat g_{t+i},\quad i=1,\dots, s,\label{eq:GMFCtEx}\\
&W_{t+i} = W_{t+i-1}\hat R_{t+i},\quad i=1,\dots, s,\label{eq:GMFCtExW}
\end{align}
where $\hat g_{t+i}$ and $\hat R_{t+i}$ represent the consumption propensity at time $t+i$ and the wealth growth in the period $t+i-1$ to $t+i$. 
We require that the choice of these two variables correctly yields the amount of consumption and investment at optimality, i.e.,
\begin{align}\label{eq:GMRationalEx}
& \hat g_{t+i} = \dfrac{C_{t+i}^\star}{W_{t+i}^\star},\; i=1,\dots, s, \\
& \hat R_{t+i} = \dfrac{W_{t+i}^\star}{W_{t+i-1}^\star},\quad i=1,\dots,s,
\end{align}
where $W_{t+i}^\star$ denotes the post-optimal consumption wealth at time $t+i$.
From \eqref{eq:GMFCtEx} and \eqref{eq:GMFCtExW}, we derive that
\begin{align}
&\hat C_{t,t+s} = \left(\hat g_{t+s}\prod_{i=2}^{s}\hat R_{t+i}\right)W_{t+1}.\label{eq:GMRPformula}
\end{align}


\begin{remark} \label{remark:WealthDependenceOfConsumptionPropensity}
Observe that, although we assume that the agent expects her future selves to follow a linear consumption rule, at the current time $t$ the agent's consumption rule is still nonlinear, i.e., $\dfrac{C_t}{W_t}$ depends on $W_t$. This is because in our model, unlike the classical case without gain-loss utility, the optimal consumption propensity, i.e., the optimal percentage of wealth consumed, should depend on the current wealth $W_t$ due to the reference point $\tilde C_{t-1,t}$. Namely, if current wealth is higher, a smaller percentage of it needs to be consumed in order to reach the reference point. \\
On this aspect, our derivation differs from that in \cite{Pagel2016:JEEA}, where the author guesses (and then verifies in equilibrium) that at each period the agent consumes a fraction of her wealth which is i.i.d. and independent of calendar time $t$ and wealth $W_{t}$.
\end{remark}

The agent's expectation (after consumption and investment) at time $t-1$ of the future consumption at time $t+s$ is modeled similarly to her expectation at time $t-1$ of the consumption at the current time $t$. Specifically, $\hat C_{t-1,t+s}$ is set to be
\begin{align}\label{eq:GMFCt1Ex}
\hat C_{t-1,t+s} = C^\star_{t-1}\tilde \varepsilon_{c,t}\frac{C^\star_{t+s}}{C^\star_{t}},
\end{align}
where $\tilde \varepsilon_{c,t}$ is independent of ${\cal F}_\infty$ and identically distributed as $\dfrac{C^\star_{t}}{C^\star_{t-1}}$ conditional on ${\cal F}_{t-1}$.

%
%

\subsection{Definition of equilibrium}\label{subsec:GM_DefinitionEquilibrium}

The economy is in equilibrium with asset prices $(R_{f,t+1},S_t),\ t \geq 0$, if the following conditions hold:
\begin{enumerate}[(i)]
\item Clearing of endowment: $C_t^{\star} = \bar{C}_t$,
\item Clearing of the stock: $\alpha_t^{\star} = 1$,
\item Rational expectations on current consumption and bounded rationality for future consumption:
\begin{enumerate}
\item $\tilde{C}_{t-1,t}$ is identically distributed as $C_t^{\star}$ conditional on $\mathcal{F}_{t-1}$.
\item $\hat C_{t,t+s}$ is given by \eqref{eq:GMRPformula} and \eqref{eq:GMRationalEx}.
\item $\hat C_{t-1,t+s}$ is given by \eqref{eq:GMFCt1Ex}.
\end{enumerate}
\end{enumerate}
The first and second conditions are clearing conditions for the aggregate consumption and the stock market, respectively. The third condition summarizes our assumptions on the agent's expectations about current and future consumption from the previous subsection. 

Before proceeding with the derivation of equilibrium asset prices, we identify the state variables of the model. 
In equilibrium, whether the consumption at time $t$ is perceived as a gain or loss depends on whether $ \bar{C}_{t}= \bar{C}_{t-1}\varepsilon_{c,t}$ is larger than the reference point $\tilde{C}_{t-1,t} = C_{t-1} \tilde{\varepsilon}_{c,t}$, where $\tilde{\varepsilon}_{c,t}$ must be interpreted as an independent copy of $\varepsilon_{c,t}$. 
Therefore, we expect that the agent's marginal utility of consumption at time $t$ depends on $\varepsilon_{c,t}$, and,  consequently, that the risk-free return, the stock price dividend-ratio, and the stock return depend on $\varepsilon_{c,t}$ as well.\footnote{From \eqref{eq:StockGrossReturn}, we can expect that the stock return will also be a function of consumption and dividend growth rates in the next period, $\varepsilon_{c,t+1}$ and $\varepsilon_{d,t+1}$.} In other words, we posit that the consumption growth rate $\varepsilon_{c,t}$ is one of the fundamental state variables.

We now argue that the consumption-dividend ratio $Y_{t}$ is also one of the state variables. If the market is in equilibrium at time $t$, then we have $C_t^{\star} = \bar{C}_t$ and $\alpha_t^{\star} = 1$, and the optimal post-consumption wealth $W_{t}^\star$ must be equal to the stock price, i.e.,
$W_{t}^\star-C^\star_t = S_t$.
The latter yields that
\begin{align}
W_t^\star= S_t + C^\star_t =S_t +  \bar{C}_t.\label{eq:GMEquiWealth}
\end{align}
Now, recalling \eqref{eq:GMRPformula}, for $s=1,2,\dots,$ we derive
\begin{align*}
\hat C_{t,t+s}& = \left(\hat g_{t+s}\prod_{i=2}^{s}\hat R_{t+i}\right)W_{t+1} = \left(\hat g_{t+s}\prod_{i=2}^{s}\hat R_{t+i}\right)W_{t+1}^\star \frac{W_{t+1}}{W_{t+1}^\star} \\[0.2cm]
& = C_{t+s}^\star \frac{W_{t+1}}{W_{t+1}^\star} =  \bar{C}_{t+s} \frac{W_{t+1}}{W_{t+1}^\star} = \frac{ \bar{C}_{t+s}}{ \bar{C}_{t+1}}\cdot \frac{ \bar{C}_{t+1}}{W_{t+1}^\star} W_{t+1},
\end{align*}
where the third equality is due to \eqref{eq:GMRationalEx} and the fourth equality follows from the market clearing conditions. Combining the above with \eqref{eq:GMEquiWealth}, we derive
\begin{align}
\hat C_{t,t+s} = \frac{ \bar{C}_{t+s}}{ \bar{C}_{t+1}}\cdot \frac{ \bar{C}_{t+1}/D_{t+1}}{S_{t+1}/D_{t+1} +  \bar{C}_{t+1}/D_{t+1}} W_{t+1}.\label{eq:GMFCtExEqui}
\end{align}
On the other hand, still in market equilibrium,
\begin{align}
\hat C_{t-1,t+s} = C^\star_{t-1}\tilde \varepsilon_{c,t}\frac{C^\star_{t+s}}{C^\star_{t}} =  \bar{C}_{t-1}\tilde \varepsilon_{c,t}\frac{ \bar{C}_{t+s}}{ \bar{C}_{t}}=  \bar{C}_{t-1}\tilde \varepsilon_{c,t} \, \varepsilon_{c,t+1}\frac{ \bar{C}_{t+s}}{ \bar{C}_{t+1}},\label{eq:GMFCt1ExEqui}
\end{align}
where $\tilde \varepsilon_{c,t}$ is independent of ${\cal F}_\infty$ and identically distributed as $\varepsilon_{c,t}$. As a result, defining
\begin{align}
\tilde U_{t,t+s}:=\mu \left(m(\hat C_{t,t+s})-m(\hat{C}_{t-1,t+s})\right),\label{eq:Uts}
\end{align}
we have
\begin{align*}
\tilde U_{t,t+s}= \mu \Bigg(m\left( \frac{ \bar{C}_{t+s}}{ \bar{C}_{t+1}}\cdot \frac{ \bar{C}_{t+1}/D_{t+1}}{S_{t+1}/D_{t+1} +  \bar{C}_{t+1}/D_{t+1}} W_{t+1}\right)-m\left( \bar{C}_{t-1} \, \tilde \varepsilon_{c,t} \,\varepsilon_{c,t+1}\,\frac{ \bar{C}_{t+s}}{ \bar{C}_{t+1}}\right)\Bigg),
\end{align*}
with
\begin{equation*}
\hat C_{t,t+s}|_{W_{t+1}=W_{t+1}^\star} =  \frac{ \bar{C}_{t+s}}{ \bar{C}_{t+1}}\cdot \frac{ \bar{C}_{t+1}/D_{t+1}}{S_{t+1}/D_{t+1} +  \bar{C}_{t+1}/D_{t+1}} W_{t+1}^\star =  \bar{C}_{t+s} =  \bar{C}_{t-1}\varepsilon_{c,t}\varepsilon_{c,t+1}\,\frac{ \bar{C}_{t+s}}{ \bar{C}_{t+1}}.\\[0.1cm]
\end{equation*}
Then, straightforward calculations yield 
\begin{equation}\label{eq:GMFutureGLUDerivative}
\begin{split}
\frac{\partial \tilde U_{t,t+s}}{\partial W_{t+1}}\Big|_{W_{t+1}=W_{t+1}^\star} = m'( \bar{C}_{t+s})\frac{ \bar{C}_{t+s}}{ \bar{C}_{t+1}}\cdot \frac{ \bar{C}_{t+1}/D_{t+1}}{S_{t+1}/D_{t+1} +  \bar{C}_{t+1}/D_{t+1}}\left(\mathbf 1_{\varepsilon_{c,t}>\tilde \varepsilon_{c,t}} + \lambda \mathbf 1_{\varepsilon_{c,t}<\tilde \varepsilon_{c,t}}\right).
\end{split}
\end{equation}

At time $t$, the consumption $ \bar{C}_t$ is perceived by the agent as a gain (loss) compared to her time-$t-1$ expectation about consumption at time $t$,  $\tilde C_{t-1,t}$, if the realization of the consumption growth rate $\varepsilon_{c,t}$ is larger (smaller) than the agent's expectation about $\tilde \varepsilon_{c,t}$. 
Therefore, the marginal utility for consumption depends on the probability that the consumption growth rate is greater than what the agent expects, which is $\mathbb{P}( \tilde \varepsilon_{c,t}\le \varepsilon_{c,t})=F_{\varepsilon_c}(\varepsilon_{c,t})$, where $F_{\varepsilon_c}(\cdot)$ denotes the cumulative distribution function of the consumption growth rate. 

We can then compute the conditional expectation of $\frac{\partial \tilde U_{t,t+s}}{\partial W_{t+1}}\Big|_{W_{t+1}=W_{t+1}^\star}$ in \eqref{eq:GMFutureGLUDerivative} as follows:
\begin{align}\label{eq:GMFutureGLUDerivativeEx}
\mathbb{E}_t\left[\frac{\partial \tilde U_{t,t+s}}{\partial W_{t+1}}\Big|_{W_{t+1}=W_{t+1}^\star}\right]  = \, & \mathbb{E}\left[m'\left(\frac{ \bar{C}_{t+s}}{ \bar{C}_{t+1}}\right)\frac{ \bar{C}_{t+s}}{ \bar{C}_{t+1}}\right]\Big(F_{\varepsilon_c}(\varepsilon_{c,t}) + \lambda (1-F_{\varepsilon_c}(\varepsilon_{c,t}))\Big)\notag \\
& \times \mathbb{E}_t\left[m'\left(\frac{ \bar{C}_{t+1}}{ \bar{C}_{t}}\right)\frac{ \bar{C}_{t+1}/D_{t+1}}{S_{t+1}/D_{t+1} + \bar{C}_{t+1}/D_{t+1}}\right]m'( \bar{C}_{t}),
\end{align}
where for the equality we use the fact that $m'$ is homogeneous of degree $\theta$, that $\dfrac{ \bar{C}_{t+i}}{ \bar{C}_{t+i-1}}$ is independent of ${\cal F}_{t+i-1}$ and $\tilde \varepsilon_{c,t}$ is independent of ${\cal F}_\infty$.

The right-hand side of Eq. \eqref{eq:GMFutureGLUDerivativeEx} shows that the quantity $\mathbb{E}_t\left[\frac{\partial \tilde U_{t,t+s}}{\partial W_{t+1}}\Big|_{W_{t+1}=W_{t+1}^\star}\right]$, which measures the (expected) sensitivity of the prospective gain-loss utility with respect to the wealth in the next period, depends on the aggregate consumption-dividend ratio $Y_{t}$.

These arguments lead us to conjecture that the risk-free return and the price-dividend ratio are functions of the consumption-dividend ratio and the consumption growth rate at time $t$, i.e., that $R_{f,t+1} = R_f\left(\varepsilon_{c,t},Y_{t}\right)$ and $\dfrac{S_t}{D_t} = g\left(\varepsilon_{c,t},Y_{t}\right)$, for some functions $R_f$ and $g$. In addition, making use of the relation in \eqref{eq:DivGrowthFromConDivRatio}, the stock return
\begin{align*}
R_{S,t+1} &= \frac{S_{t+1}+D_{t+1}}{S_t} = \frac{S_{t+1}/D_{t+1}+1}{S_t/D_t} \cdot \frac{D_{t+1}}{D_t} = \frac{g\left(\varepsilon_{c,t+1}, Y_{t+1}\right)+1}{g\left(\varepsilon_{c,t},Y_{t}\right)}\,\varepsilon_{d,t+1}\\
& = \frac{g\left(\varepsilon_{c,t+1},Y_{t+1}\right)+1}{g\left(\varepsilon_{c,t},Y_{t}\right)} \,  \varepsilon_{c,t+1} \frac{Y_{t}}{Y_{t+1}},
\end{align*}
can be written as a function of $\varepsilon_{c,t}, \, \varepsilon_{c,t+1}, Y_{t}$, and $Y_{t+1}$.


To summarize, we can let $\varepsilon_{c,t}$ and $Y_t$ be the state variables of the model, and assume that (i) $\varepsilon_{c,t}$ is independent of ${\cal F}_{t-1}$ and has the same distribution for every $t$, and that (ii) $\{Y_t\}_{t\in\mathbb{N}}$ is a Markov chain. 

\subsection{Equilibrium prices}
The following theorem provides equilibrium asset prices in semi-closed form for Model I.

\begin{theorem}\label{GM_thm}
Assume that the following growth condition holds:
\begin{equation*}
\beta \mathbb{E}\bigg[\varepsilon_{c,t+1}^{1-\theta}\bigg] < 1.
\end{equation*}
Then, the equilibrium risk-free return is
\begin{eqnarray}
\begin{aligned}
&R_{f,t+1} =\Big(1+ bF_{\varepsilon_c}(\varepsilon_{c,t})+ b \lambda (1-F_{\varepsilon_c}(\varepsilon_{c,t}))\Big)\Bigg(\gamma \Big(F_{\varepsilon_c}(\varepsilon_{c,t}) + \lambda(1-F_{\varepsilon_c}(\varepsilon_{c,t})) \Big) \\
& \quad \times \beta\mathbb{E}_t \left[  \varepsilon_{c,t+1}^{-\theta}Y_{t+1}\frac{1+ bF_{\varepsilon_c}(\varepsilon_{c,t+1})+ b \lambda (1-F_{\varepsilon_c}(\varepsilon_{c,t+1}))}{Y_{t+1}\Big(1+ bF_{\varepsilon_c}(\varepsilon_{c,t+1})+  b \lambda(1-F_{\varepsilon_c}(\varepsilon_{c,t+1}))\Big)\left(1-\beta \mathbb{E}
\left[\varepsilon_{c,t+1}^{1-\theta} \right]\right)+h(\varepsilon_{c,t+1},Y_{t+1})} \right] \Bigg. \\
& \Bigg. \quad + \beta\mathbb{E} \left[  \varepsilon_{c,t+1}^{-\theta}   \Big( 1+ bF_{\varepsilon_c}(\varepsilon_{c,t+1})+ b \lambda (1-F_{\varepsilon_c}(\varepsilon_{c,t+1})) \Big) \right]\Bigg)^{-1}\label{eq:GM_EquiRiskFree},
\end{aligned}
\end{eqnarray}
and the equilibrium stock price-dividend ratio is
\begin{eqnarray}
\begin{aligned}
\frac{S_{t}}{D_{t}}= \frac{h(\varepsilon_{c,t},Y_{t})}{\Big(1+ bF_{\varepsilon_c}(\varepsilon_{c,t})+ b \lambda (1-F_{\varepsilon_c}(\varepsilon_{c,t}))\Big)\left(1-\beta \mathbb{E}
\left[\varepsilon_{c,t+1}^{1-\theta} \right]\right)},
\end{aligned}
\end{eqnarray}
where $h$ is the solution to the following integral equation:
\begin{equation} 
\begin{split}
&h(\varepsilon_{c,t},Y_t) = \gamma\Big(F_{\varepsilon_c}(\varepsilon_{c,t}) + \lambda(1-F_{\varepsilon_c}(\varepsilon_{c,t}))\Big)  \\
& \quad \times \beta \mathbb{E}_t
\left[ \varepsilon_{c,t+1}^{1-\theta}Y_{t}\frac{\Big(1+ bF_{\varepsilon_c}(\varepsilon_{c,t+1})+ b \lambda (1-F_{\varepsilon_c}(\varepsilon_{c,t+1}))\Big)\left(1-\beta \mathbb{E}
\left[\varepsilon_{c,t+1}^{1-\theta} \right]\right)+h(\varepsilon_{c,t+1},Y_{t+1})}{Y_{t+1}\Big(1+ bF_{\varepsilon_c}(\varepsilon_{c,t+1})+ b \lambda (1-F_{\varepsilon_c}(\varepsilon_{c,t+1}))\Big)\left(1-\beta \mathbb{E}
\left[\varepsilon_{c,t+1}^{1-\theta} \right]\right)+h(\varepsilon_{c,t+1},Y_{t+1})}  \right]  \\[0.2cm]
& \quad +   \beta\mathbb{E}_t
\left[ \varepsilon_{c,t+1}^{1-\theta}\dfrac{Y_{t}}{Y_{t+1}}
\bigg( h(\varepsilon_{c,t+1},Y_{t+1})+ \Big(1+ bF_{\varepsilon_c}(\varepsilon_{c,t+1})+ b \lambda (1-F_{\varepsilon_c}(\varepsilon_{c,t+1}))\Big)\left(1-\beta \mathbb{E}
\left[\varepsilon_{c,t+1}^{1-\theta} \right]\right)\bigg)  \right].\label{eq:GM_hFun}
\end{split}
\end{equation}
In addition, the stock return is
\begin{equation} \label{eq:GM_EquiStockReturn}
\begin{split}
\hspace{-0.5cm} R_{S,t+1} = \, & \varepsilon_{c,t+1}\dfrac{Y_{t}}{Y_{t+1}}\left(\dfrac{1+ bF_{\varepsilon_c}(\varepsilon_{c,t})+ b \lambda(1-F_{\varepsilon_c}(\varepsilon_{c,t}))}{h\left(\varepsilon_{c,t},Y_{t}\right)} \right) \\
& \times \left( \dfrac{h\left(\varepsilon_{c,t+1},Y_{t+1}\right)}{1+ bF_{\varepsilon_c}(\varepsilon_{c,t+1})+ b \lambda(1-F_{\varepsilon_c}(\varepsilon_{c,t+1}))} + 1-\beta \mathbb{E}
	\left[\varepsilon_{c,t+1}^{1-\theta} \right] \right) .\\[0.2cm]
\end{split}
\end{equation}

\end{theorem}

\begin{corollary} \label{SDF_general_cor}
In the setting of Theorem \ref{GM_thm}, the stochastic discount factor $M_{t+1}$ is given by
\begin{equation*}
\begin{split}
M_{t+1} & = \frac{\beta}{1+ bF_{\varepsilon_c}(\varepsilon_{c,t})+ b \lambda(1-F_{\varepsilon_c}(\varepsilon_{c,t})) } \Bigg ( \varepsilon_{c,t+1}^{-\theta} \big( 1+ bF_{\varepsilon_c}(\varepsilon_{c,t+1})+ b \lambda(1-F_{\varepsilon_c}(\varepsilon_{c,t+1})) \big) \Bigg. \\[0.2cm]
& \quad \Bigg. +  \gamma\beta    \varepsilon_{c,t+1}^{-\theta}Y_{t+1}\dfrac{\big(F_{\varepsilon_c}(\varepsilon_{c,t}) + \lambda(1-F_{\varepsilon_c}(\varepsilon_{c,t})) \big)\big(1+ bF_{\varepsilon_c}(\varepsilon_{c,t+1})+ b \lambda (1-F_{\varepsilon_c}(\varepsilon_{c,t+1}))\big)}{Y_{t+1}\Big(1+ bF_{\varepsilon_c}(\varepsilon_{c,t+1})+  b \lambda(1-F_{\varepsilon_c}(\varepsilon_{c,t+1}))\Big)\left(1-\beta \mathbb{E}
\left[\varepsilon_{c,t+1}^{1-\theta} \right]\right)+h(\varepsilon_{c,t+1},Y_{t+1})} \Bigg).
\end{split}
\end{equation*}
\end{corollary}

A numerical analysis of the results in Theorem \ref{GM_thm} is postponed to Section \ref{sec:Numerics}, while the formal proof is provided in Appendix \ref{subsec:GM_thm_proof}. Therein, we show that equilibrium asset prices derive from the following Euler equations for optimality of the agent's portfolio selection problem:
\vspace{0.2cm}
\begin{equation*}
\begin{split}
& 1 =  \beta \mathbb{E}_t\left[\left(\dfrac{\bar{C}_{t+1}}{\bar{C}_{t}}\right)^{-\theta}\dfrac{\big(1+ bF_{\varepsilon_c}(\varepsilon_{c,t+1})+ b \lambda(1-F_{\varepsilon_c}(\varepsilon_{c,t+1}))\big)}{1+ bF_{\varepsilon_c}(\varepsilon_{c,t})+ b \lambda(1-F_{\varepsilon_c}(\varepsilon_{c,t}))}R_{S,t+1}\right]\\
& \quad + \frac{\gamma\beta}{1-\beta \mathbb{E}[\varepsilon_{c,t+1}^{1-\theta}]} \cdot \dfrac{F_{\varepsilon_c}(\varepsilon_{c,t}) + \lambda(1-F_{\varepsilon_c}(\varepsilon_{c,t}))}{1+ bF_{\varepsilon_c}(\varepsilon_{c,t})+ b \lambda(1-F_{\varepsilon_c}(\varepsilon_{c,t}))} \, \mathbb{E}_t\left[\left(\dfrac{\bar{C}_{t+1}}{\bar{C}_{t}}\right)^{-\theta}\frac{Y_{t+1}}{S_{t+1}/D_{t+1} + Y_{t+1}}R_{S,t+1}\right],\\[0.2cm]
&1 = \beta \mathbb{E}_t\left[\left(\dfrac{\bar{C}_{t+1}}{\bar{C}_{t}}\right)^{-\theta}\dfrac{\big(1+ bF_{\varepsilon_c}(\varepsilon_{c,t+1})+ b \lambda(1-F_{\varepsilon_c}(\varepsilon_{c,t+1}))\big)}{1+ bF_{\varepsilon_c}(\varepsilon_{c,t})+ b \lambda(1-F_{\varepsilon_c}(\varepsilon_{c,t}))}\right]R_{f,t+1} \\
&\quad +\frac{\gamma\beta}{1-\beta \mathbb{E}[\varepsilon_{c,t+1}^{1-\theta}]} \cdot \dfrac{F_{\varepsilon_c}(\varepsilon_{c,t}) + \lambda(1-F_{\varepsilon_c}(\varepsilon_{c,t}))}{1+ bF_{\varepsilon_c}(\varepsilon_{c,t})+ b \lambda(1-F_{\varepsilon_c}(\varepsilon_{c,t}))} \, \mathbb{E}_t\left[\left(\dfrac{\bar{C}_{t+1}}{\bar{C}_{t}}\right)^{-\theta}\frac{Y_{t+1}}{S_{t+1}/D_{t+1} + Y_{t+1}}\right]R_{f,t+1}. \\[0.2em]
\end{split}
\end{equation*}
Compared to the classical asset pricing model with power utility over consumption, the first term on the right-hand side of both equations is relatively standard, as it measures the trade-off between the return of each asset and the marginal utility of consumption, in this case with a supplementary multiplicative factor that depends on the gain-loss component. The second term is instead new and stems out of the prospective gain-loss utility. The impact of this term is mostly determined by the expected growth of aggregate consumption and the covariance between the marginal utility of consumption and (a function of) the state variable $Y_{t+1}$. 

Moving on, now we state a technical proposition showing that the function $h(\varepsilon_{c,t}, Y_{t})$ in \eqref{eq:GM_hFun} has a unique fixed point. Notably, a direct implication of Proposition \ref{prop:FixedPoint} is the existence of a solution for the equilibrium asset pricing problem in Model I. 
\begin{proposition}\label{prop:FixedPoint}
Assume that $\varepsilon_{c,t},$ for $ t=0,1,\dots$, are i.i.d. Then, the right-hand side of \eqref{eq:GM_hFun} is a contraction mapping in the space of functions that are larger than or equal to a lower bound $\underline{h}$, where $\underline{h}$ is implicitly defined by
\begin{equation*}
\begin{split}
\mathbb{E}_{t}
\left[\Bigg( \beta \varepsilon_{c,t+1}^{1-\theta}\dfrac{Y_{t}}{Y_{t+1}} \frac{\gamma \lambda \Big(1+ bF_{\varepsilon_c}(\varepsilon_{c,t+1})+  b \lambda(1-F_{\varepsilon_c}(\varepsilon_{c,t+1}))\Big)\left(1-\beta \mathbb{E}
\left[\varepsilon_{c,t+1}^{1-\theta} \right]\right)}{\Big(\big(1+ bF_{\varepsilon_c}(\varepsilon_{c,t+1})+ b \lambda (1-F_{\varepsilon_c}(\varepsilon_{c,t+1}))\big)\left(1-\beta \mathbb{E}
\left[\varepsilon_{c,t+1}^{1-\theta} \right]\right)+\underline{h}\Big)^2} 	+ \beta \varepsilon_{c,t+1}^{1-\theta} \dfrac{Y_{t}}{Y_{t+1}}\Bigg)^2 \right] = 1 .\\[1cm]
\end{split}
\end{equation*}
\end{proposition}

\subsection{Consumption-wealth ratio} 
We end our presentation of Model I with some comments on the equilibrium consumption-wealth ratio. As shown in Appendix \ref{appendix:CWratio_generalmodel}, this is given by
\begin{equation*}
\begin{split}
\dfrac{\bar{C}_{t}}{W_t}   = & \left( \rule{0cm}{1cm} 1 +\beta \mathbb{E}_t \left[ \rule{0cm}{1cm}\varepsilon_{c,t+1}^{1-\theta} \dfrac{\big( 1+ bF_{\varepsilon_c}(\varepsilon_{c,t+1})+ b \lambda(1-F_{\varepsilon_c}(\varepsilon_{c,t+1})) \big)}{\big( 1+ bF_{\varepsilon_c}(\varepsilon_{c,t})+ b \lambda(1-F_{\varepsilon_c}(\varepsilon_{c,t
})) \big)} \right] \right. \\
& \left. \hspace{2cm} + \beta \mathbb{E}_t \left[  \rule{0cm}{1cm}  \varepsilon_{c,t+1}^{1-\theta}  \dfrac{ h(\varepsilon_{c,t+2}, Y_{t+2})}{\dfrac{\bar{C}_{t+1}}{D_{t+1}}\big( 1+ bF_{\varepsilon_c}(\varepsilon_{c,t})+ b \lambda(1-F_{\varepsilon_c}(\varepsilon_{c,t})) \big) \Big(1-\beta \mathbb{E} \left[\varepsilon_{c,t+2}^{-\theta}\right]\Big)} \right]\right)^{-1}. \\
\end{split}
\end{equation*}
Essentially, the first expectation corresponds to the expected gain-loss utility of consumption at time $t+1$ divided by the contemporaneous gain-loss utility, whereas the second term gauges the expected marginal utility of consumption times the benefits of holding the risky stock, again scaled by the contemporaneous gain-loss component.

In \cite{Pagel2016:JEEA}, Proposition 2, the author shows that, for all realizations of the consumption growth rate, under some conditions on the preference parameters the consumption-wealth ratio is lower than in the traditional model (that is, without gain-loss utility and with consumption \textit{equal to} dividends).\footnote{In the traditional model, the consumption-wealth ratio has a simple representation in terms of the discounted stream of future consumption utilities: $ \dfrac{\bar{C}_{t}}{W_{t}} = \left(1+\mathbb{E}_{t}\left[\sum_{\tau=1}^{\infty}\beta^{\tau}\left(\dfrac{\bar{C}_{t+\tau}}{\bar{C}_{t}}   \right)^{1-\theta}\right] \right)^{-1}$.} In our setting, where aggregate consumption is \textit{different from} dividends, in the absence of gain-loss utility ($b = \gamma = 0$) the equilibrium consumption-wealth ratio simplifies to
\begin{equation*} 
\dfrac{\bar{C}_{t}}{W_t} \Big\vert_{b = \gamma = 0}   = \left(1 +\beta \mathbb{E}_t \left[\varepsilon_{c,t+1}^{1-\theta}\left(1 + \dfrac{ \beta\mathbb{E}_{t+1} \bigg[\varepsilon_{c,t+2}^{-\theta}\varepsilon_{d,t+2} \bigg]}{\dfrac{\bar{C}_{t+1}}{D_{t+1}} \Big(1-\beta \mathbb{E}_{t+1} \left[\varepsilon_{c,t+2}^{-\theta}\varepsilon_{d,t+2}\right]\Big)} \right) \right]\right)^{-1}.
\end{equation*}
Despite the simplification, it does not appear possibile to provide a coincise condition in terms of the preference parameters so that $\dfrac{\bar{C}_{t}}{W_t} \leq \dfrac{\bar{C}_{t}}{W_t} \Big\vert_{b = \gamma =0}  $ for all values of $\varepsilon_{c,t}$. Whether the equilibrium consumption-wealth ratio is lower than in the case without gain-loss utility depends in a non-trivial way on the value of the consumption growth rate.

\section{Model II: contemporaneous gain-loss utility only}
\label{sec:BaselineModel}
We consider a special case of the general Model I in which the agent's preferences do not include prospective gain-loss utility ($\gamma = 0$). 

\subsection{Preferences} \label{subsec:BaselineModel_Preferences}
In this case, the cumulative utility at time $t$  for a consumption stream $\{C_s\}_{s\ge t}$ is given by
\begin{equation}
V_t\left(\{C_s\}_{s\ge t}  \mid \{\tilde{C}_{s-1,s}\}_{s\ge t}\right) = \mathbb{E}_t\left[\sum_{s=t}^{\infty} \beta^{s-t} U_s(C_s)\right],
\end{equation}
where the instantaneous utility of consuming $C_s$ at time $s$ with reference point $\tilde C_{s-1,s}$ is given by
\begin{eqnarray}\label{baseline_utility}
U_s(C_s) = m(C_s)+ b \, \mathbb{E}_s \left[ \mu \left( m(C_s) - m(\tilde{C}_{s-1,s}) \right) \right].
\end{eqnarray}


\subsection{Definition of equilibrium}\label{subsec:Baseline_Equi}

The economy is in equilibrium with asset prices $(R_{f,t+1},S_t),\ t \geq 0$, if the following conditions hold:
\begin{enumerate}[(i)]
\item Clearing of endowment: $C_t^{\star} = \bar{C}_t$,
\item Clearing of the stock: $\alpha_t^{\star} = 1$,
\item Rational expectations on current consumption: $\tilde{C}_{t-1,t}$ is identically distributed as $C_t^{\star}$, conditional on $\mathcal{F}_{t-1}$.
\end{enumerate}

\subsection{Equilibrium prices}

Under the same market setting described in Subsection \ref{subsec:TheMarket}, the next theorem provides equilibrium asset prices in explicit form for Model II.

\begin{theorem}\label{Baseline_thm}
Assume that the following growth condition holds:
\begin{equation} \label{growthcondition}
\beta \mathbb{E}_{t} \left[\varepsilon_{c,t+1}^{-\theta}\varepsilon_{d,t+1} \right] < 1.
\end{equation}
Then, there exists an equilibrium with risk-free return
\begin{equation} \label{eq:RiskFreeReturn_BaselineModel}
R_{f,t+1} =  \frac{1+ bF_{\varepsilon_c}(\varepsilon_{c,t})+ b \lambda(1-F_{\varepsilon_c}(\varepsilon_{c,t})) }{\beta\mathbb{E}\left[  \varepsilon_{c,t+1}^{-\theta} \big( 1+ bF_{\varepsilon_c}(\varepsilon_{c,t+1})+ b \lambda(1-F_{\varepsilon_c}(\varepsilon_{c,t+1})) \big) \right]},
\end{equation}
and stock price-dividend ratio
\begin{equation} \label{eq:StockPriceDividendRatio_BaselineModel}
\frac{S_{t}}{D_{t}} =  \frac{ \beta\mathbb{E}_{t} \bigg[\varepsilon_{c,t+1}^{-\theta}\varepsilon_{d,t+1}\Big(1+ bF_{\varepsilon_c}(\varepsilon_{c,t+1})+ b \lambda(1-F_{\varepsilon_c}(\varepsilon_{c,t+1}))\Big)\bigg]}{\Big(1+ bF_{\varepsilon_c}(\varepsilon_{c,t})+ b \lambda(1-F_{\varepsilon_c}(\varepsilon_{c,t}))\Big) \Big(1-\beta \mathbb{E}_{t} \left[\varepsilon_{c,t+1}^{-\theta}\varepsilon_{d,t+1}\right]\Big)}.
\end{equation} 
In addition, the stock return is
\begin{equation} \label{eq:StockReturn:BaselineModel}
\begin{split}
 & \quad R_{S,t+1} = \, \Big(1+ bF_{\varepsilon_c}(\varepsilon_{c,t})+ b \lambda(1-F_{\varepsilon_c}(\varepsilon_{c,t}))\Big)\varepsilon_{d,t+1} \\ 
&  \times  \left(\frac{1-\beta \mathbb{E}_{t} \Big[\varepsilon_{c,t+1}^{-\theta}\varepsilon_{d,t+1}\Big]}{ \beta \mathbb{E}_{t} \left[\varepsilon_{c,t+1}^{-\theta}\varepsilon_{d,t+1}\Big(1+ bF_{\varepsilon_c}(\varepsilon_{c,t+1})+ b \lambda(1-F_{\varepsilon_c}(\varepsilon_{c,t+1}))\Big)\right]}  +\frac{1 }{1+ bF_{\varepsilon_c}(\varepsilon_{c,t+1})+ b \lambda(1-F_{\varepsilon_c}(\varepsilon_{c,t+1}))} \right),
\end{split}
\end{equation}
and the conditional risk premium is
\begin{equation} \label{eq:RiskPremium_BaselineModel}
\begin{split}
& \quad \mathbb{E}_t\big[R_{S,t+1}\big]-R_{f,t+1} = \Big(1+ bF_{\varepsilon_c}(\varepsilon_{c,t})+ b \lambda(1-F_{\varepsilon_c}(\varepsilon_{c,t}))\Big)\\
\hspace{-1.2cm}& \times \left(\rule{0cm}{1cm} \mathbb{E}_{t}\left[\frac{\varepsilon_{d,t+1}}{1+ bF_{\varepsilon_c}(\varepsilon_{c,t+1})+ b \lambda(1-F_{\varepsilon_c}(\varepsilon_{c,t+1}))}\right] +\frac{\mathbb{E}_{t}\left[\varepsilon_{d,t+1}\right]\Big(1-\beta \mathbb{E}_{t}\left[\varepsilon_{c,t+1}^{-\theta}\varepsilon_{d,t+1}\right]\Big)}{\beta \mathbb{E}_{t}\left[\varepsilon_{c,t+1}^{-\theta}\varepsilon_{d,t+1}\Big(1+ bF_{\varepsilon_c}(\varepsilon_{c,t+1})+ b \lambda(1-F_{\varepsilon_c}(\varepsilon_{c,t+1})) \Big)\right]} \right.\\
\hspace{-1.2cm}& \left. \quad - \frac{1}{\beta\mathbb{E}_{t}\left[ \varepsilon_{c,t+1}^{-\theta}\Big(1+ bF_{\varepsilon_c}(\varepsilon_{c,t+1})+ b \lambda(1-F_{\varepsilon_c}(\varepsilon_{c,t+1}))\Big)\right]}   \rule{0cm}{1cm}\right).
\end{split}
\end{equation}
\end{theorem}


\begin{corollary}\label{SDF_baseline_cor}
In the setting of Theorem \ref{Baseline_thm}, the stochastic discount factor $M_{t+1}$ is given by
\begin{equation*} \label{eq:SDF_baseline}
M_{t+1} = \frac{ \beta  \varepsilon_{c,t+1}^{-\theta} \big( 1+ bF_{\varepsilon_c}(\varepsilon_{c,t+1})+ b \lambda(1-F_{\varepsilon_c}(\varepsilon_{c,t+1})) \big) }{1+ bF_{\varepsilon_c}(\varepsilon_{c,t})+ b \lambda(1-F_{\varepsilon_c}(\varepsilon_{c,t})) }.
\end{equation*}
\end{corollary}

\vspace{0.3cm}

The reader might find it helpful to draw a connection between the above formulas and the results for Model I (\mbox{Theorem \ref{GM_thm}}). In the latter case, equilibrium prices are shown to depend on the function $ h(\varepsilon_{c,t},Y_t)$ defined in \eqref{eq:GM_hFun}, which contains an involved term depending on $\gamma$. When $\gamma$ is set equal to 0, \eqref{eq:GM_hFun} simplifies to 
\begin{equation*} 
\begin{split}
&h(\varepsilon_{c,t},Y_t) - \beta\mathbb{E}_t
\left[ \varepsilon_{c,t+1}^{1-\theta}\dfrac{Y_{t}}{Y_{t+1}}
 h(\varepsilon_{c,t+1},Y_{t+1})  \right] \\
& \hspace{2cm} = \beta\mathbb{E}_{t}
\left[ \varepsilon_{c,t+1}^{1-\theta}\dfrac{Y_{t}}{Y_{t+1}}
\Big(1+ bF_{\varepsilon_c}(\varepsilon_{c,t+1})+ b \lambda (1-F_{\varepsilon_c}(\varepsilon_{c,t+1}))\Big) \right]\left(1-\beta \mathbb{E}
\left[\varepsilon_{c,t+1}^{1-\theta} \right]\right), \\[0.2cm]
\end{split}
\end{equation*}
and equilibrium prices under Model II can then be retrieved for
\begin{equation*}
h(\varepsilon_{c,t},Y_t)  = \dfrac{\beta\mathbb{E}_{t}
\left[ \varepsilon_{c,t+1}^{-\theta}\varepsilon_{d,t+1}
\Big(1+ bF_{\varepsilon_c}(\varepsilon_{c,t+1})+ b \lambda (1-F_{\varepsilon_c}(\varepsilon_{c,t+1}))\Big) \right]\left(1-\beta \mathbb{E}
\left[\varepsilon_{c,t+1}^{1-\theta} \right]\right)}{1-\beta\mathbb{E}_{t}
\left[ \varepsilon_{c,t+1}^{-\theta}\varepsilon_{d,t+1}  \right]}. \\[0.2cm]
\end{equation*}

The additional tractability of Model II also allows us to derive some intuitive results that showcase more directly the effect on equilibrium asset prices of some of the ingredients of the model. We start with a proposition that shows the comparative statics for the risk-free rate, the price-dividend ratio, and the risky return with respect to the consumption growth rate.

\begin{proposition}\label{prop:BaselineModel_ComparativeConsGrowth}
In the setting of Theorem \ref{Baseline_thm}, it holds that:
\begin{enumerate}
\item[(i)] The risk-free return $R_{f,t+1}$ is decreasing in $\varepsilon_{c,t}$.
\item[(ii)] The price-dividend ratio $\dfrac{S_{t}}{D_{t}}$ is increasing in $\varepsilon_{c,t}$.
\item[(iii)] If $\varepsilon_{d,t+1} > 0$, the risky return $R_{S,t+1}$ is decreasing in $\varepsilon_{c,t}$; otherwise, $R_{S,t+1}$ is increasing in $\varepsilon_{c,t}$.
\end{enumerate}
\end{proposition}

Intuitively, because of loss aversion, the marginal utility of consumption is larger when consumption is in the loss region than when it is in the gain region. Consequently, \emph{ceteris paribus}, the agent is less willing to consume more today when $\varepsilon_{c,t}$ is larger. 
In equilibrium, this yields that the risk-free rate (respectively, stock-price dividend ratio) is decreasing (resp., increasing) in the current consumption growth rate $\varepsilon_{c,t}$. 
For the risky return, this relation is conditioned by the value of the dividend growth rate $\varepsilon_{d,t+1}$. 

The next proposition shows the comparative statics for the risk-free rate and the price-dividend ratio with respect to the parameters of the gain-loss utility function.
\begin{proposition}\label{prop:BaselineModel_ComparativeParametersGainLoss}
In the setting of Theorem \ref{Baseline_thm}, it holds that:
\begin{enumerate}
\item[(i)]If
\begin{equation} \label{eq:RiskFreeRate_comparative_statics_threshold}
F_{\varepsilon_c}(\varepsilon_{c,t}) > \frac{\mathbb{E}[\varepsilon_{c,t+1}^{-\theta} F_{\varepsilon_c}(\varepsilon_{c,t+1})]}{\mathbb{E}[\varepsilon_{c,t+1}^{-\theta}]},
\end{equation}
then the risk-free return rate $R_{f,t+1}$ is decreasing in $\lambda$ and $b$; otherwise, $R_{f,t+1}$ is increasing in $\lambda$ and $b$.
\item[(ii)] If
\begin{equation} \label{eq:StockPriceDivRatio_comparative_statics_threshold}
F_{\varepsilon_c}(\varepsilon_{c,t}) > \frac{\mathbb{E}_{t}[\varepsilon_{c,t+1}^{-\theta}\varepsilon_{d,t+1} F_{\varepsilon_c}(\varepsilon_{c,t+1})]}{\mathbb{E}_{t}[\varepsilon_{c,t+1}^{-\theta}\varepsilon_{d,t+1}]},
\end{equation}
then the price-dividend ratio $\dfrac{S_t}{D_t}$ is increasing in $\lambda$ and $b$; otherwise, $\dfrac{S_t}{D_t}$ is decreasing in $\lambda$ and $b$.
\end{enumerate}

\noindent [Note: The comparative statics for the risky return turned out to depend on a much less succint condition, therefore we opted to not spell it out. A numerical investigation is provided in Section \ref{sec:Numerics}.]

\end{proposition}

Proposition \ref{prop:BaselineModel_ComparativeParametersGainLoss}(i) shows that whether a higher degree of loss aversion $\lambda$ leads to a higher or lower risk-free return depends on the growth rate of the consumption in the current period. The intuition for this result is as follows. An increase in $\lambda$ increases the marginal utility of consumption both at current time $t$ and in the next period $t+1$. The magnitude of the increment depends on the likelihood of consumption to be in the loss region: the more likely that consumption is in the loss region, the stronger the effect of loss aversion on the marginal utility, thus the increment will be larger. As a result, if the consumption growth rate in the current period is high (according to the condition in \eqref{eq:RiskFreeRate_comparative_statics_threshold}), the consumption in the current period is more likely to be in the gain region, so the increment in the marginal utility of consumption in the current period - due to a higher loss aversion degree - is smaller than the increment in the marginal utility of consumption in the next period. Consequently, the agent is more willing to save or invest and consume in the future, and this drives down the equilibrium risk-free rate. By the same mechanism, as per Proposition \ref{prop:BaselineModel_ComparativeParametersGainLoss}(ii), an increase in $\lambda$ drives up the equilibrium stock-price dividend ratio if $F_{\varepsilon_{c}}(\varepsilon_{c,t})$ is higher than the threshold in \eqref{eq:StockPriceDivRatio_comparative_statics_threshold}.

A similar reasoning can explain the effect of the weighting factor $b$ on $R_{f,t+1}$ and $\dfrac{S_t}{D_t}$. In this case, because an increase in $b$ represents a more significant weight that the agent attributes to the gain-loss component of the utility function, her willingness to consume in the current period will depend again on the probability that the actual consumption is larger than the expected consumption at time $t$. If such probability is higher than the threshold in \eqref{eq:RiskFreeRate_comparative_statics_threshold} (respectively, \eqref{eq:StockPriceDivRatio_comparative_statics_threshold}), the increment in the marginal utility of consumption in the current period is smaller than the increment in the marginal utility of consumption in the next period. Consequently, this induces less consumption at time $t$ and again a lower (resp., higher) equilibrium risk-free rate (resp., stock-price dividend ratio).


\section{Numerical results} \label{sec:Numerics}

In this section we conduct a numerical analysis of the results for Models I and II. In particular, we will first study the effect on equilibrium prices of the consumption growth rate and consumption-dividend ratio, then provide a sensitivity analysis of the mean and standard deviation of prices with respect to a number of parameters of interest.\footnote{
Here we refer to the \emph{unconditional} mean and standard deviation:  $\mathbb{E}\left[R_{f,t+1}\right], \sqrt{\big.\mbox{Var}\left[R_{f,t+1}\right]\big.}$ for the risk-free rate; $\mathbb{E}\left[\dfrac{S_{t}}{D_{t}}\right], \sqrt{\mbox{Var}\left[\dfrac{S_{t}}{D_{t}}\right]}$ for the price-dividend ratio;  ${\mathbb{E}\big[R_{S,t+1}-R_{f,t+1}\big]}, \sqrt{\mbox{Var}\big[R_{S,t+1}-R_{f,t+1}\big]}$ for the risk premium. }

\paragraph{Distributional assumptions and choice of parameters.} Recall our modeling assumptions from Subsection \ref{subsec:TheMarket} on the consumption growth rate and the consumption-dividend ratio, respectively:
\begin{equation*}
\begin{split}
\log(\varepsilon_{c,t+1}) & = \mu_{c} + \sigma_{c}z_{t+1}, \quad z_{t} \sim \mathcal{N}(0,1) \mbox{ i.i.d.},\\[0.2em]
\log(Y_{t+1}) & = (1-\varphi)\kappa + \varphi\log(Y_{t}) +  \sigma_{y}\epsilon_{t+1}, \quad \epsilon_{t} \sim \mathcal{N}(0,1) \mbox{ i.i.d.}
\end{split}
\end{equation*}
Note that the conditional distribution of $\log(Y_{t+1})$ reads as $$\log(Y_{t+1}) \, \vert \, \log(Y_{t})\sim \mathcal{N}\Big( (1-\varphi)\kappa +  \varphi \log(Y_{t}), \sigma_{y}^{2}\Big).$$ In addition, if $\vert \varphi \vert < 1$, the process $\log(Y_{t})$ is stationary and has unconditional distribution $$\log(Y_{t}) \sim \mathcal{N}\left( \kappa, \frac{\sigma_{y}^{2}}{1-\varphi^{2}}\right).$$
In order to find the distribution of the dividend growth rate (which is needed for Model II), we use the relation \eqref{eq:DivGrowthFromConDivRatio}. As the model for $\varepsilon_{c,t+1}$ has been specified above, we only need to find the distribution of the ratio $\dfrac{Y_{t+1}}{Y_t}$. On this account, we can write
\begin{equation*}
\log\left(\frac{Y_{t+1}}{Y_{t}}\right) = \varphi \log\left(\dfrac{Y_{t}}{Y_{t-1}}\right) +  \sigma_{y} \left(\epsilon_{t+1} - \epsilon_{t}\right), \vspace{0.1cm}
\end{equation*}
which is easily shown to have conditional distribution 
$$\log\left(\dfrac{Y_{t+1}}{Y_{t}}\right) \, \bigg\vert \, \log\left(\dfrac{Y_{t}}{Y_{t-1}}\right) \sim \mathcal{N}\Bigg( \varphi \log\left(\dfrac{Y_{t}}{Y_{t-1}}\right), 2\sigma_{y}^{2}\Bigg), \vspace{0.1cm}$$
and unconditional distribution
$$\log\left(\dfrac{Y_{t+1}}{Y_{t}}\right) \sim \mathcal{N}\Bigg( 0,  \dfrac{2(1-\varphi)\sigma_{y}^{2}}{1-\varphi^{2}}\Bigg). \vspace{0.3cm}$$
Table \ref{Table:CalibrationConsumptionGrowth_ConsumptionDividend} shows our maximum likelihood estimation for the parameters $\mu_{c},\sigma_{c},\varphi, \kappa, \sigma_{y}$ obtained using annual data for the period 1929-2022 from the U.S. Department of Commerce, Bureau of Economic Analysis. This set of parameters leads to an average consumption growth rate of 1.061, with standard deviation 0.056, and a long-term average consumption-dividend ratio of 17.826, with standard deviation 6.59.

\begin{table}[h!]
\centering
\begin{tabular}{ll}
\hline \hline
$\mu_{c}$ \qquad & 0.058 \\
$\sigma_{c}$ \qquad & 0.053  \\
$\varphi$ \qquad & 0.961\\
$\kappa$ \qquad & 2.816\\
$\sigma_{y}$ \qquad &  0.099\\
\hline \hline
\end{tabular}
\caption{Parameter values for the consumption growth and the consumption-dividend ratio.  }\label{Table:CalibrationConsumptionGrowth_ConsumptionDividend}
\end{table}

To conclude on parametrization choices, throughout this section we keep fixed the discounting parameter $\beta = 0.98$ and the risk aversion $\theta = 4$. Other relevant parameters, namely the degree of loss aversion $\lambda$, the weight of the contemporaneous gain-loss utility component $b$, and the weight of the prospective gain-loss utility component $\gamma$, will be specified case by case.

\paragraph{Equilibrium prices: the effect of consumption growth and consumption-dividend ratio.}

\begin{figure}[t!]
\begin{minipage}[t]{0.33\textwidth}
	\centering
\subfloat[Risk-free rate]{\includegraphics[width=1\textwidth]{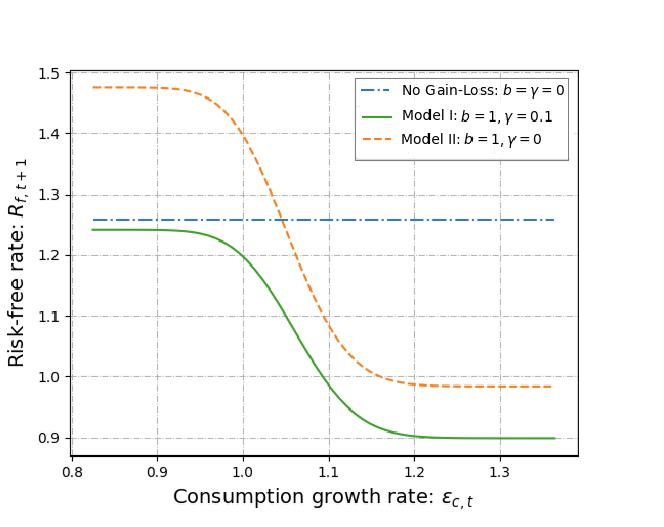}}
\subfloat[Stock price-dividend ratio]{\includegraphics[width=1\textwidth]{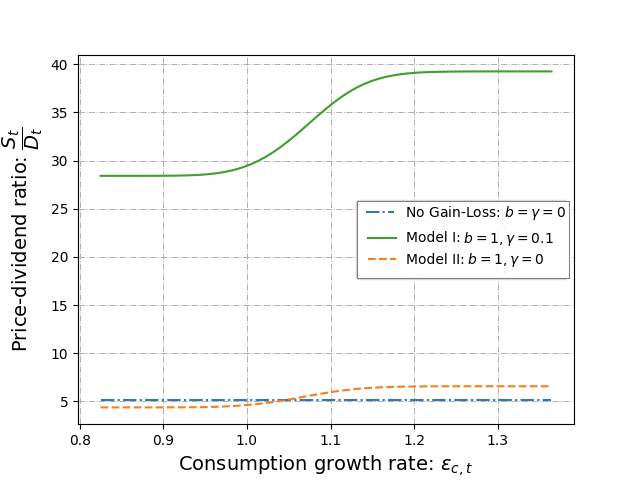}}
\subfloat[Equity risk premium]{\includegraphics[width=1\textwidth]{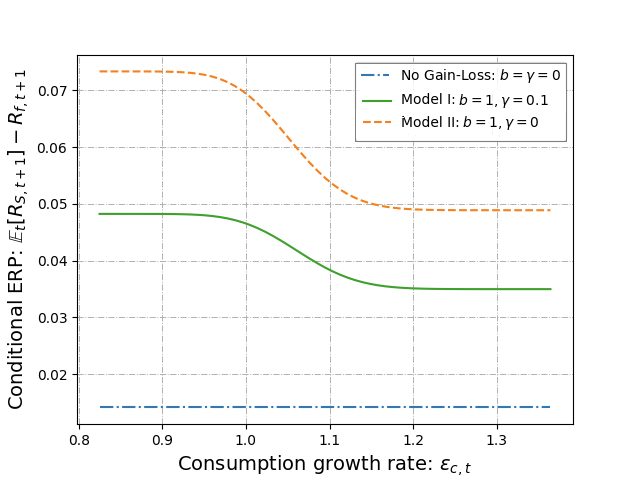}}
\end{minipage}
\caption{Equilibrium prices in Models I and II and without the gain-loss component, as function of the consumption growth rate. For Model I, we choose an average consumption-dividend ratio $Y_{t} = 21.07$. Other parameters are set as follows: $\theta = 4, \lambda =2,\beta = 0.98$. }
\label{fig:EquilibriumPricesAsFunctionOfConsgrowth}
\end{figure} 

\begin{figure}[t!]
\begin{minipage}[t]{0.33\textwidth}
	\centering
\subfloat[Risk-free rate]{\includegraphics[width=1\textwidth]{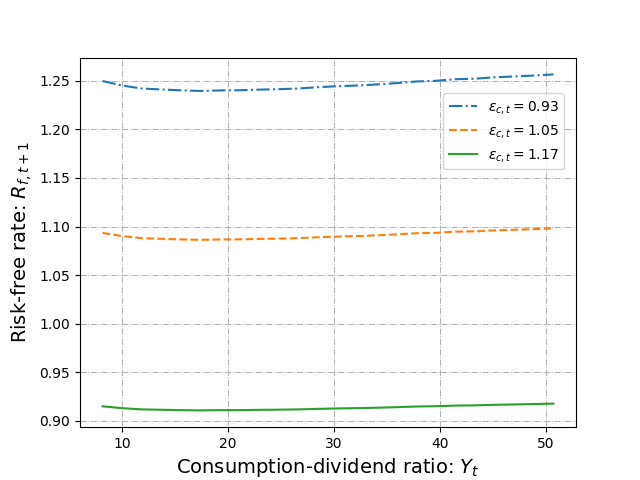}}
\subfloat[Stock price-dividend ratio]{\includegraphics[width=1\textwidth]{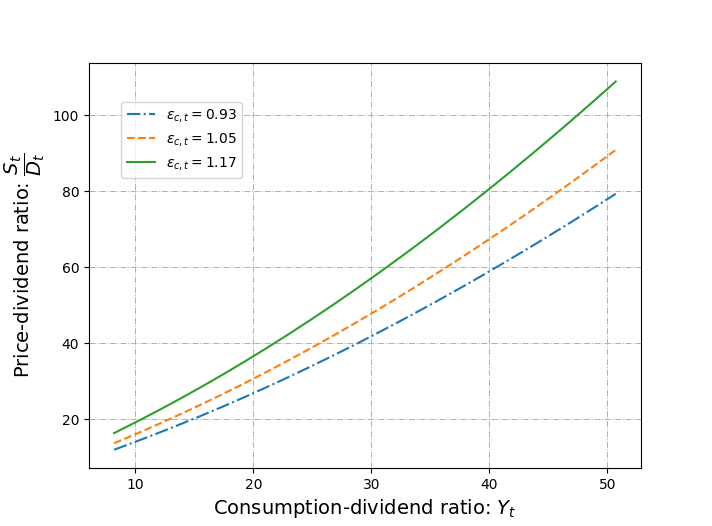}}
\subfloat[Equity risk premium]{\includegraphics[width=1\textwidth]{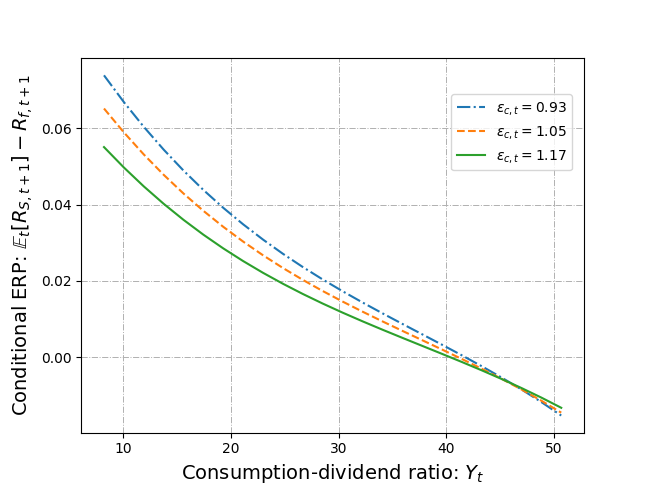}}
\end{minipage}
\caption{Equilibrium prices in Model I as function of the consumption-dividend ratio, for different values of the consumption growth rate. Other parameters are set as follows: $\theta = 4,b = 1, \lambda = 2, \gamma = 0.1, \beta = 0.98$. }
\label{fig:EquilibriumPricesAsFunctionOfConsDivratio}
\end{figure} 

Figure \ref{fig:EquilibriumPricesAsFunctionOfConsgrowth} shows the risk-free rate (Subfigure \ref{fig:EquilibriumPricesAsFunctionOfConsgrowth}(a)), the stock price-dividend ratio (Subfigure \ref{fig:EquilibriumPricesAsFunctionOfConsgrowth}(b)), and the conditional equity risk premium (Subfigure \ref{fig:EquilibriumPricesAsFunctionOfConsgrowth}(c)) as function of $\varepsilon_{c,t}$ in Model I (solid green curve, $b=1, \gamma=0.1$), Model II (dashed red curve, $b=1, \gamma = 0$), and in a model without gain-loss utility (dash-dotted blue curve, $b=\gamma=0$).  
For Model I, in which equilibrium prices depend also on the current consumption-dividend ratio, we fix $Y_{t}$ as the historical average based on our data: 21.07. Similarly, Figure \ref{fig:EquilibriumPricesAsFunctionOfConsDivratio} shows the risk-free rate (Subfigure \ref{fig:EquilibriumPricesAsFunctionOfConsDivratio}(a)), the stock-price dividend ratio (Subfigure \ref{fig:EquilibriumPricesAsFunctionOfConsDivratio}(b)), and the conditional equity risk premium (Subfigure \ref{fig:EquilibriumPricesAsFunctionOfConsDivratio}(c)\footnote{To reduce numerical noise, the curves in Figure \ref{fig:EquilibriumPricesAsFunctionOfConsDivratio}(c) have been smoothed using a B-spline interpolation.}) in Model I as function of $Y_{t}$, for different values of $\varepsilon_{c,t}$: 1.05 (dashed red curve), 0.93 (dash-dotted blue curve), 1.17 (solid green curve), where the values correspond to the historical average of the consumption growth rate, minus/plus one standard deviation. In both figures, $\lambda =2$. 

We previously argued that, in the presence of loss aversion, the marginal utility of consumption at time $t$ is higher when $\varepsilon_{c,t}$ is low, as the agent is more likely to experience a gain. Therefore, she will consume more in the current period, while saving less and investing less in the risky stock. This idea motivates why the risk-free rate (respectively, stock price-dividend ratio) is decreasing (increasing) in the current consumption growth, as we know from Proposition \ref{prop:BaselineModel_ComparativeConsGrowth}. In addition, seeing that the agents' reference point in the future is built upon the current consumption, consuming more now will create higher expectations for consumption in the next period. Hence, since the agent is loss averse, she will expect to receive a higher equity premium in the next period. We quantify these effects in Table \ref{Table:EquilibriumPricesConsumptionGrowth}, where we report average prices in Model II for different values of the consumption growth rate. For instance, the risk-free rate is about 48\% larger when the consumption growth rate is one s.d. below its mean than when it is one s.d. above its mean, whereas the price-dividend ratio is 32\% smaller (-2.083 in absolute terms) and the equity premium is 2.4\% larger.
\begin{table}
\centering
\begin{tabular}{lccc}
\hline \hline
\qquad & $\varepsilon_{c,t} = 0.93$ & $\varepsilon_{c,t} = 1.05$ & $\varepsilon_{c,t} = 1.17$ \\
\hline
$R_{f,t+1}$ & 1.47  & 1.24  & 0.99\\[0.2em]
$\dfrac{S_{t}}{D_{t}}$ & 4.41 & 5.224 & 6.493   \\[0.2em]
$\mathbb{E}_{t}\left[R_{s,t+1}\right] - R_{f,t+1}$  & 0.073  & 0.061 & 0.049 \\[0.2em]
\hline \hline
\end{tabular}
\caption{Equilibrium prices in Model II corresponding to an average (center) consumption growth rate and average minus-plus (left and right, respectively) one standard deviation.
Other parameters are set as follows: $\theta = 4, b = 1, \lambda = 2, \beta = 0.98$.}
\label{Table:EquilibriumPricesConsumptionGrowth}
\end{table}

As shown in the figures, the mechanisms described above also apply to Model I - for all that on average there we obtain lower risk-free rates and equity premia, but higher price-dividend ratios. Also, we note that in Model I the effect of the aggregate consumption-dividend ratio on the equilibrium risk-free rate is negligible, while the effect on the price-dividend ratio and the equity risk premium is in line with what we have seen for the consumption growth.

Finally, it is worthwile to point out that, given aggregate consumption growth as a procyclical economic indicator, these figures suggest that both our models qualitatively conform with well-known stylized facts: procyclical variation in the price-dividend ratio (\cite{FamaFrench1989:JFE}), countercyclical variation in the conditional expected equity premium (\cite{CampbellShiller1988:JF, CampbellShiller1988:RFS}, \cite{FamaFrench1989:JFE}), and countercyclical variation in the conditional volatility of the equity premium (\cite{BollerslevChouKroner1992:JE}). On the other hand, the model fails to capture procyclical variation of risk-free rates (\cite{Fama1990:JME}).

\paragraph{Equilibrium prices: the effect of preference parameters.}

We now assess the sensitivity of equilibrium prices with respect to the preference parameters. In doing so, we compare our results with empirical estimates based on annualized data from the Kenneth R. French Data Library for the period 1929-2022.\footnote{The empirical values reported in the tables should be treated as indicative, for they might change significantly based on the choice of, e.g., the time period, the frequency of evaluation (quarterly, biannually, annually, etc.), or the risk-free rate; see \cite{WelchGoyal2008:RFS}.}

\begin{table}[t!]
\centering
\begin{tabular}{lccccc}
\hline \hline
\qquad &  $b = 1, \lambda=2$ & $ b = 1.5, \lambda = 2$ &  $b = 1, \lambda=3$ & $ b = 1.5, \lambda = 3$ & Empirical value\\
\hline
Risk-free rate &&&&& \\
\quad Mean  & 1.072  & 1.081 &   1.054  & 1.064 & 1.033   \\
\quad Standard deviation  &  0.098 &  0.12 & 0.169 &  0.198&  0.031 \\
Price-dividend ratio &&&&& \\
\quad Mean &  26.508  & 19.48  & 32.71 &  24.736  & 33.28    \\
\quad Standard deviation &  10.38 & 6.931 &  14.253 &  10.432 & 18.28   \\
Equity risk premium &&&&& \\
\quad Mean &  0.048  &  0.06 & 0.083 & 0.092 & 0.082 \\
\quad Standard deviation & 0.165  & 0.188 & 0.254  & 0.293   & 0.201 \\
\hline \hline
\end{tabular}
\caption{Equilibrium prices in Model I as function of the loss aversion parameter $\lambda$ and the gain-loss utility weight $b$. Other parameters as set as follows: $\theta = 4, \gamma = 0.1, \beta = 0.98$. Empirical values are based on annualized data from the Kenneth R. French Data Library for the period 1929-2022. \\}
\label{Table:GeneralPricesFunction_lambda_and_b_theta4}
\end{table}

\begin{table}[h!]
\centering
\begin{tabular}{lccccc}
\hline \hline
\qquad & $\gamma=0$ &  $\gamma=0.05$ & $\gamma = 0.1$  & $\gamma = 0.5$ & Empirical value\\
\hline
Risk-free rate &&&&& \\
\quad Mean & 1.229   & 1.113   & 1.072  & 1.041 & 1.03 \\
\quad Standard deviation & 0.142  & 0.109 & 0.098  &  0.091 & 0.031  \\
Price-dividend ratio &&&&& \\
\quad Mean & 5.375 &  13.327 &  26.508  & 167.78 & 33.28    \\
\quad Standard deviation & 0.785  & 3.787  & 10.38    & 77.435 &  18.28   \\
Equity risk premium &&&&& \\
\quad Mean & 0.058 & 0.049  & 0.048   & 0.044 & 0.082 \\
\quad Standard deviation & 0.223 &  0.177  & 0.165  & 0.161  & 0.201 \\
\hline \hline
\end{tabular}
\caption{Equilibrium prices in Model I as function of the prospective gain-loss utility weight $\gamma$. Other parameters as set as follows: $b = 1, \lambda = 2, \theta = 4, \beta = 0.98$. Empirical values are based on annualized data from the Kenneth R. French Data Library for the period 1929-2022. \\}
\label{Table:GeneralPricesFunction_gamma_theta4}
\end{table}

\begin{table}[t!]
\centering
\begin{tabular}{lcccccc}
\hline \hline
\qquad & $\lambda=1$ & $\lambda = 1.5$ & $\lambda = 2$ & $\lambda = 2.5$ &  $\lambda=3$ & Empirical value \\
\hline
Risk-free rate &&&&&& \\
\quad Mean &  1.258   & 1.242   &  1.229  & 1.219& 1.21 &  1.033  \\
\quad Standard deviation & 0.001   & 0.08 & 0.142 &  0.192 &  0.233 & 0.031  \\
Price-dividend ratio &&&&&& \\
\quad Mean & 5.173   &  5.245 & 5.335  & 5.442  &  5.549  & 33.28   \\
\quad Standard deviation & 0.005 & 0.338  &  0.626 &  0.884 & 1.118  & 18.28   \\
Equity risk premium &&&&&& \\
\quad Mean &  0.014  &  0.037 &  0.061   &  0.085 & 0.11 & 0.082  \\
\quad Standard deviation &  0.144  & 0.188  &  0.237   & 0.285  &  0.332 &   0.201 \\
\hline \hline
\end{tabular}
\caption{Equilibrium prices in Model II as function of the loss aversion parameter $\lambda$. Other parameters as set as follows: $\theta = 4, b = 1, \beta = 0.98$. Empirical values are based on annualized data from the Kenneth R. French Data Library for the period 1929-2022.
\\}
\label{Table:BaselinePricesFunction_lambda_theta4}
\end{table}

\begin{table}[t!]
\centering
\begin{tabular}{lccccc}
\hline \hline
\qquad & $b=0$ & $b=0.5$  & $b=1$  &  $b=1.5$ & Empirical value\\
\hline
Risk-free rate &&&&& \\
\quad Mean &  1.258   &  1.237  &   1.229   &  1.224 &  1.033 \\
\quad Standard deviation & 0.001  & 0.102 &  0.142 &  0.163 & 0.031  \\
Price-dividend ratio &&&&& \\
\quad Mean & 5.165 &    5.275   &  5.375  &  5.38 & 33.28   \\
\quad Standard deviation & 0.004 &  0.438  &   0.785 & 0.737  &  18.28  \\
Equity risk premium &&&&& \\
\quad Mean &  0.014  &  0.045    &  0.058  & 0.07   & 0.082 \\
\quad Standard deviation &  0.144  &   0.204  &  0.223  & 0.256 &  0.201  \\
\hline \hline
\end{tabular}
\caption{Equilibrium prices in Model II as function of the gain-loss utility weight $b$.  Other parameters are set as follows: $\theta = 4,\lambda = 2, \beta = 0.98$. Empirical values are based on annualized data from the Kenneth R. French Data Library for the period 1929-2022. \\}
\label{Table:BaselinePricesFunction_b_theta4}
\end{table}

In Table \ref{Table:GeneralPricesFunction_lambda_and_b_theta4}, we display the unconditional mean and standard deviation of the risk-free rate, the price-dividend ratio, and the equity risk premium in Model I for different combinations of $b = 1,1.5$ and $\lambda = 2,3$, with $\gamma = 0.1$. This is complemented with Table \ref{Table:GeneralPricesFunction_gamma_theta4} showing the unconditional mean and standard deviation of the risk-free rate, the price-dividend ratio, and the equity risk premium in Model I for different values of $\gamma =$ 0, 0.01, 0.05, 0.1, 0.5.\footnote{The first column is a corner case where Model I reduces to the special case of Model II. Ergo, values are equivalent with column 3 in Table \ref{Table:BaselinePricesFunction_lambda_theta4} and column 4 in Table \ref{Table:BaselinePricesFunction_b_theta4}.} Here we fix $b=1, \lambda=2$.

Equilibrium prices in Model I are strongly sensitive to the choice of the prospective gain-loss utility weight. Under the current parametrization, for larger values of $\gamma$ the agent decides more and more to postpone consumption and invest heavily in the present, which makes the risk-free rate and the equity risk premium (respectively, the price-dividend ratio) decreasing (increasing) in $\gamma$. 

For different levels of $\gamma$, our model generates average estimates for the risk-free rate \textit{or} the price-dividend ratio that closely match the data. For instance, setting $\gamma = 0.5$, the model produces a $4.1 \%$ risk-free rate, though at the expense of a overly high price-dividend ratio, around 168. Oppositely, reducing $\gamma$ to 0.1 yields a price-dividend ratio of $\approx$ 26.5, with a risk-free rate of 7.2\%. In both cases, for $b=1$ and $\lambda=2$, the ERP appears to be slightly lower than the empirical value, so one would need to increase these parameters if mostly oriented in producing a higher ERP (as shown in Table \ref{Table:GeneralPricesFunction_lambda_and_b_theta4}). 

Surprisingly, here we also observe that the effect of $b$ on the risk-free rate and the price-dividend ratio is opposite to the effect of $\lambda$, despite both parameters reflecting loss aversion. One possible explanation for this is that in Model I, $b$ does not only correspond to the weight of contemporaneous loss aversion but also reflects a relative weight between current and prospective gain-loss utility. We will see below, in fact, that in absence of prospective gain-loss utility $b$ and $\lambda$ have similar effects on asset prices.

We now turn our attention to Model II. Table \ref{Table:BaselinePricesFunction_lambda_theta4} presents the unconditional mean and standard deviation of the risk-free rate, the price-dividend ratio and the equity risk premium in Model II for different values of $\lambda =$ 1, 1.5, 2, 2.5, 3. Here we fix $b$ as 1 and $\theta$ as 4. Similarly, Table \ref{Table:BaselinePricesFunction_b_theta4} presents the unconditional mean and standard deviation of the risk-free rate, the price-dividend ratio and the equity risk premium in Model II for different values of $b =$ 0, 0.5, 1, 1.5. The value of $\lambda$ here is 2.

In Model II, increases in $\lambda$ and $b$ imply that the agent is more often disposed to postpone consumption as she aims to avoid encountering current losses - in the first case because losses become more painful, in the second for she attributes a heavier weight to the contemporaneous gain-loss component of the utility function. This habit tends to decrease (respectively, increase) the equilibrium risk-free rate (price-dividend ratio). Yet, within the range of parameters used in our experiments, this effect is not severe. On the contrary, the impact of $\lambda$ and $b$ on the equity risk premium is more pronounced, hinting at the notion that more loss averse agents will demand a higher equity premium. In particular, for $\lambda = 2.5$ and $b = 1$, our model can generate an 8.5\% equity risk premium, with a reasonable standard deviation of 28.5\%. 

While the model is able to produce a sizeable equity risk premium, for the chosen risk aversion ($\theta = 4$) it also produces high equilibrium risk-free rates and low price-dividend ratios. Besides introducing prospective gain-loss utility, one strategy to obtain lower risk-free rates and higher price-dividend ratios here could be to reduce the value of $\theta$, yet at the expense of lower equity premia. For instance, for $\theta = 1.2$, $b=1$, $\lambda = 3$, Model II produces a risk-free rate of 7.8\% (s.d. 21\%), a price-dividend ratio of 38.4 (s.d. 7.75), and an equity risk premium of 7.1\% (s.d. 31.6\%). In order to produce an average risk-free rate of, say, 3\%, one would need to set $\theta$ to about 0.2. However, values of $\theta < 1$ likely violate the growth condition in \eqref{growthcondition} and result in negative price-dividend ratios. For this reason, and given that the role of risk aversion has been amply documented already in the literature, we decided not to provide an in-depth sensitivity analysis and to keep it constant throughout the section. For reference, additional results for $\theta = 2$ and different values of the prospective gain-loss weight $\gamma$ are provided in Appendix \ref{appendix:AdditionalNumerics}.

As a final note, we point out that Tables \ref{Table:BaselinePricesFunction_lambda_theta4} and \ref{Table:BaselinePricesFunction_b_theta4} confirm the results in Proposition \ref{prop:BaselineModel_ComparativeParametersGainLoss} on the monotonicity of equilibrium risk-free rate and price-dividend ratio as functions of $\lambda$ and $b$. 

\section{Conclusions} \label{sec:Conclusions}
We studied a discrete-time consumption-based capital asset pricing model under the reference-dependent preferences proposed by \cite{KoszegiRabin2006:QJE, KoszegiRabin2007:AER,KoszegiRabin2009:AER}. More precisely, we considered an endowment economy populated by a representative agent who, on top of utility from current consumption, derives utility from gains and losses in consumption with respect to a reference point. This reference point is assumed to be forward-looking - for it is based on the agent's expectations about future consumption - and stochastic - for it depends on the future consumption growth rate. 

Compared to the work of \cite{Pagel2016:JEEA}, our model separates between consumption and dividend streams and allows for optimal strategies that are wealth-dependent. These differences lead to some technical challenges that we addressed in two models. First, we considered a general model (Model I) in which the agent derives both contemporaneous and prospective gain-loss utility, that is, utility from the difference between current consumption and previously held expectations about current consumption, and utility from the difference between intertemporal beliefs about future consumption, respectively. A semi-closed form solution for equilibrium asset prices was derived in this case.  Then, we considered a special case (Model II) in which the agent derives contemporaneous gain-loss utility only, obtaining equilibrium asset prices in closed form. By analysing the effect on equilibrium asset prices of several preference parameters, we found that, with plausible values of risk aversion and loss aversion, our models can generate equity premia that are comparable to empirical estimates.
Furthermore, both the general model and the special case are consistent with some well-known empirical
facts, namely procyclical variation in the price-dividend ratio and
countercyclical variation in the conditional expected equity premium and in the conditional volatility of the equity premium.
Finally, we found that prospective gain-loss utility is necessary for the model to predict reasonable values of the price-dividend ratio.


\bibliography{LongTitles,BibFile_AssetPricingGainLoss}

\appendix

\section{Proofs and additional theoretical results}

\subsection{Proof of Theorem \ref{GM_thm}} \label{subsec:GM_thm_proof}

Based on the discussion in Subsection \ref{subsec:GM_DefinitionEquilibrium}, we can consider Markovian controls for our problem, i.e., we can consider the consumption $C_t$ and  the percentage of investment in the stock $\alpha_t$ to be functions of current-time state variables $(W_t, \bar{C}_t,Y_t,\varepsilon_{c,t})$. In a compact form, we will write $C_t = C(W_t, \bar{C}_t,Y_t,\varepsilon_{c,t})$ and $\alpha_t = \alpha(W_t, \bar{C}_t,Y_t,\varepsilon_{c,t})$, for some functions $C$ and $\alpha$. In addition, we denote by $f(W_t, \bar{C}_t, Y_t, \varepsilon_{c,t})$  the optimal value of the agent's portfolio selection problem. 

The following dynamic programming equation holds:
\begin{eqnarray}
f(W_t, \bar{C}_t, Y_t, \varepsilon_{c,t}) = \underset{C_t,\alpha_t}{\max}\ \ \tilde{U}_t+\beta \mathbb{E}_t \left[ f(W_{t+1}, \bar{C}_{t+1}, Y_{t+1}, \varepsilon_{c,t+1}) \right],\label{eq:GM_DP}
\end{eqnarray}
where
\begin{eqnarray}\label{GM_preferenceSim}
\begin{aligned}
\tilde{U}_t&= m(C_t)+b \mathbb{E}_t \left[ \mu \left( m( C_t) - m(\tilde {C}_{t-1,t}) \right) \right] + \gamma \sum_{s=1}^{\infty}\beta^s  \mathbb{E}_t \left[\tilde U_{t,t+s}\right]\\
& = U_t(C_t)+ \gamma \sum_{s=1}^{\infty}\beta^s  \mathbb{E}_t \left[\tilde U_{t,t+s}\right],
\end{aligned}
\end{eqnarray}
from the definition of $\tilde U_{t,t+s}$ in \eqref{eq:Uts}. 

The first-order conditions for optimality are then as follows:
\begin{align}
\frac{\partial \tilde U_t}{\partial C_t}\Big|_{C_t=C_t^\star,\alpha_t = \alpha_t^\star} + \beta \frac{\partial \mathbb{E}_t \left[ f(W_{t+1}, \bar{C}_{t+1}, D_{t+1}, \varepsilon_{c,t+1}) \right]}{\partial C_t}\Big|_{C_t=C_t^\star,\alpha_t = \alpha_t^\star}=0,\label{eq:GMFOCCt}\\[0.2cm]
\frac{\partial \tilde U_t}{\partial \alpha_t}\Big|_{C_t=C_t^\star,\alpha_t = \alpha_t^\star} + \beta \frac{\partial \mathbb{E}_t \left[ f(W_{t+1}, \bar{C}_{t+1}, D_{t+1}, \varepsilon_{c,t+1}) \right]}{\partial \alpha_t}\Big|_{C_t=C_t^\star,\alpha_t = \alpha_t^\star}=0. \label{eq:GMFOCalphat}
\end{align}
We now proceed to compute the terms in \eqref{eq:GMFOCCt} and \eqref{eq:GMFOCalphat}. First, note that $\tilde{U}_t$ depends on both $C_t$ and $W_{t+1}$, and that in equilibrium $C_t^\star =  \bar{C}_t$ and $\alpha_t^\star = 1$, where $C^\star_t$ and $\alpha^\star_t$ denote the optimal consumption and allocation in the stock, respectively. Thus, we can compute
\begin{align}
\frac{\partial U_t(C_t)}{\partial C_t}\Big|_{C_t = C_t^\star} = \Big(1+ bF_{\varepsilon_c}(\varepsilon_{c,t})+ b \lambda(1-F_{\varepsilon_c}(\varepsilon_{c,t}))\Big)m'( \bar{C}_t).\label{eq:BM_UtCt}
\end{align}
Moreover, considering that
\begin{align}
&\frac{\partial W_{t+1}}{\partial C_t}\Big|_{C_t=C_t^\star,\alpha_t = \alpha_t^\star} = -R_{S,t+1},\label{eq:Wt1DeriveCt}\\
&\frac{\partial W_{t+1}}{\partial \alpha_t}\Big|_{C_t=C_t^\star,\alpha_t = \alpha_t^\star} = \left(W_{t}- \bar{C}_{t}\right)\left(R_{S,t+1}-R_{f,t+1}\right),\label{eq:Wt1Derivealphat}
\end{align}
together with \eqref{eq:GMFutureGLUDerivative}, it follows that
\begin{align*}
\frac{\partial \mathbb{E}_t[\tilde U_{t,t+s}]}{\partial C_t}\Big|_{C_t=C_t^\star,\alpha_t = \alpha_t^\star} &= - \bar{C}_t^{-\theta}\mathbb{E}\left[\left(\frac{ \bar{C}_{t+s}}{ \bar{C}_{t+1}}\right)^{1-\theta}\right]\big(F_{\varepsilon_c}(\varepsilon_{c,t}) + \lambda(1-F_{\varepsilon_c}(\varepsilon_{c,t})) \big)\\
&\quad \times \mathbb{E}_t\left[\varepsilon_{c,t+1}^{-\theta} \frac{Y_{t+1}}{S_{t+1}/D_{t+1} + Y_{t+1}}R_{S,t+1}\right],\\[0.2cm]
\frac{\partial \mathbb{E}_t[\tilde U_{t,t+s}]}{\partial \alpha_t}\Big|_{C_t=C_t^\star,\alpha_t = \alpha_t^\star} &=  \bar{C}_t^{-\theta}\mathbb{E}\left[\left(\frac{ \bar{C}_{t+s}}{ \bar{C}_{t+1}}\right)^{1-\theta}\right]\big(F_{\varepsilon_c}(\varepsilon_{c,t}) + \lambda(1-F_{\varepsilon_c}(\varepsilon_{c,t}))\big)\\
&\quad \times \mathbb{E}_t\left[\varepsilon_{c,t+1}^{-\theta} \frac{Y_{t+1}}{S_{t+1}/D_{t+1} + Y_{t+1}}\left(W_{t}- \bar{C}_{t}\right)(R_{S,t+1}-R_{f,t+1})\right].
\end{align*}
Given that $\mathbb{E}\left[\left(\frac{ \bar{C}_{t+s}}{ \bar{C}_{t+1}}\right)^{1-\theta}\right] = \mathbb{E}\left[\varepsilon_{c,t+1}^{1-\theta}\right]^{s-1}$, we derive
\begin{align}
&\frac{\partial \tilde U_t}{\partial C_t}\Big|_{C_t=C_t^\star,\alpha_t = \alpha_t^\star} =  \bar{C}_t^{-\theta} \Bigg( 1+ bF_{\varepsilon_c}(\varepsilon_{c,t})+ b \lambda(1-F_{\varepsilon_c}(\varepsilon_{c,t}))\notag\\
&\quad  - \frac{\gamma\beta}{1-\beta \mathbb{E}[\varepsilon_{c,t+1}^{1-\theta}]}\big(F_{\varepsilon_c}(\varepsilon_{c,t}) + \lambda(1-F_{\varepsilon_c}(\varepsilon_{c,t}))\big)\mathbb{E}_t\left[\varepsilon_{c,t+1}^{-\theta} \frac{Y_{t+1}}{S_{t+1}/D_{t+1} + Y_{t+1}}R_{S,t+1}\right]\Bigg),\label{eq:GMUtDeriveCt}\\[0.2cm]
&\frac{\partial \tilde U_t}{\partial \alpha_t}\Big|_{C_t=C_t^\star,\alpha_t = \alpha_t^\star} =  \bar{C}_t^{-\theta}\frac{\gamma\beta}{1-\beta \mathbb{E}[\varepsilon_{c,t+1}^{1-\theta}]}\big(F_{\varepsilon_c}(\varepsilon_{c,t}) + \lambda(1-F_{\varepsilon_c}(\varepsilon_{c,t}))\big)\notag\\
&\quad \times \mathbb{E}_t\left[\varepsilon_{c,t+1}^{-\theta} \frac{Y_{t+1}}{S_{t+1}/D_{t+1} + Y_{t+1}}\left(W_{t}- \bar{C}_{t}\right)(R_{S,t+1}-R_{f,t+1})\right].\label{eq:GMUtDerivealphat}
\end{align}
Next, \eqref{eq:Wt1DeriveCt} and \eqref{eq:Wt1Derivealphat} yield that
\begin{align}
&\frac{\partial \mathbb{E}_t \left[ f(W_{t+1}, \bar{C}_{t+1}, D_{t+1}, \varepsilon_{c,t+1}) \right]}{\partial C_t}\Big|_{C_t=C_t^\star,\alpha_t = \alpha_t^\star} \nonumber  \\
 & = -\mathbb{E}_t \left[ f_W(W_{t+1}^*, \bar{C}_{t+1}, Y_{t+1}, \varepsilon_{c,t+1})R_{S,t+1} \right],\label{eq:EfDeriveCt}\\[0.2cm]
&\frac{\partial \mathbb{E}_t \left[ f(W_{t+1}, \bar{C}_{t+1}, D_{t+1}, \varepsilon_{c,t+1}) \right]}{\partial \alpha_t}\Big|_{C_t=C_t^\star,\alpha_t = \alpha_t^\star} \nonumber \\ & = \mathbb{E}_t \left[ f_W(W_{t+1}^*, \bar{C}_{t+1}, Y_{t+1}, \varepsilon_{c,t+1})\left(W_{t}- \bar{C}_{t}\right)(R_{S,t+1}-R_{f,t+1}) \right].\label{eq:EfDerivealphat}
\end{align}
Furthermore, we have
\begin{align*}
f(W_t^\star, \bar{C}_t, Y_t,\varepsilon_{c,t}) = \tilde U_t\Big|_{C_t=C_t^\star,\alpha_t = \alpha_t^\star} + \beta \mathbb{E}_t \left[ f(W_{t+1}^*, \bar{C}_{t+1}, D_{t+1}, \varepsilon_{c,t+1}) \right].
\end{align*}
Taking the derivative with respect to $W_{t}$ on both sides of the above equality and considering \eqref{eq:GMFOCCt} and \eqref{eq:GMFOCalphat}, we derive
\begin{align*}
f_{W_t}(W_t^\star, \bar{C}_t, Y_t,\varepsilon_{c,t}) = \frac{\partial \tilde U_t}{\partial W_t}\Big|_{C_t=C_t^\star,\alpha_t = \alpha_t^\star} + \beta \frac{\partial \mathbb{E}_t \left[ f(W_{t+1}^*, \bar{C}_{t+1}, D_{t+1}, \varepsilon_{c,t+1}) \right]}{\partial W_t}\Big|_{C_t=C_t^\star,\alpha_t = \alpha_t^\star}.
\end{align*}
Noting that
\begin{align*}
&\frac{\partial \tilde U_t}{\partial W_t}\Big|_{C_t=C_t^\star,\alpha_t = \alpha_t^\star} = \frac{\partial U_t(C_t)}{\partial C_t}\Big|_{C_t=C_t^\star,\alpha_t = \alpha_t^\star}-\frac{\partial \tilde U_t}{\partial C_t}\Big|_{C_t=C_t^\star,\alpha_t = \alpha_t^\star},\\[1em]
&\frac{\partial \mathbb{E}_t \left[ f(W_{t+1}^*, \bar{C}_{t+1}, D_{t+1}, \varepsilon_{c,t+1}) \right]}{\partial W_t}\Big|_{C_t=C_t^\star,\alpha_t= \alpha_t^\star} = -\frac{\partial \mathbb{E}_t \left[ f(W_{t+1}^*, \bar{C}_{t+1}, D_{t+1}, \varepsilon_{c,t+1}) \right]}{\partial C_t}\Big|_{C_t=C_t^\star,\alpha_t = \alpha_t^\star},
\end{align*}
and recalling \eqref{eq:GMFOCCt} and \eqref{eq:BM_UtCt}, we obtain
\begin{align}
f_{W_t}(W_t^\star, \bar{C}_t, Y_t,\varepsilon_{c,t}) =  \bar{C}_t^{-\theta}\big(1+ bF_{\varepsilon_c}(\varepsilon_{c,t})+ b \lambda(1-F_{\varepsilon_c}(\varepsilon_{c,t}))\big).\label{eq:GM_fWt}
\end{align}
Moving on, using Eqs. \eqref{eq:GMUtDeriveCt}-\eqref{eq:GM_fWt} we write the following Euler equations:
\begin{align}
& \big(1+ bF_{\varepsilon_c}(\varepsilon_{c,t})+ b \lambda(1-F_{\varepsilon_c}(\varepsilon_{c,t}))\big) -\beta \mathbb{E}_t\big[\varepsilon_{c,t+1}^{-\theta}\big(1+ bF_{\varepsilon_c}(\varepsilon_{c,t+1})+ b \lambda(1-F_{\varepsilon_c}(\varepsilon_{c,t+1}))\big)R_{S,t+1}\big]\notag\\
& \quad - \frac{\gamma\beta}{1-\beta \mathbb{E}[\varepsilon_{c,t+1}^{1-\theta}]} \big(F_{\varepsilon_c}(\varepsilon_{c,t}) + \lambda(1-F_{\varepsilon_c}(\varepsilon_{c,t}))\big)\mathbb{E}_t\left[\varepsilon_{c,t+1}^{-\theta} \frac{Y_{t+1}}{S_{t+1}/D_{t+1} + Y_{t+1}}R_{S,t+1}\right]=0,\label{eq:GM_Euler1}\\[0.2cm]
& \frac{\gamma\beta}{1-\beta \mathbb{E}[\varepsilon_{c,t+1}^{1-\theta}]}\big(F_{\varepsilon_c}(\varepsilon_{c,t}) + \lambda(1-F_{\varepsilon_c}(\varepsilon_{c,t}))\big) \mathbb{E}_t\left[\varepsilon_{c,t+1}^{-\theta} \frac{Y_{t+1}}{S_{t+1}/D_{t+1} + Y_{t+1}}(R_{S,t+1}-R_{f,t+1})\right]\notag\\
& \quad +\beta \mathbb{E}_t\left[\varepsilon_{c,t+1}^{-\theta}\big(1+ bF_{\varepsilon_c}(\varepsilon_{c,t+1})+ b \lambda(1-F_{\varepsilon_c}(\varepsilon_{c,t+1})\big)(R_{S,t+1}-R_{f,t+1})\right]=0.\label{eq:GM_Euler2}
\end{align}
Combining \eqref{eq:GM_Euler1} and \eqref{eq:GM_Euler2}, we have
\begin{align}
&\bigg( \frac{\gamma\beta}{1-\beta \mathbb{E}[\varepsilon_{c,t+1}^{1-\theta}]}\big(F_{\varepsilon_c}(\varepsilon_{c,t}) + \lambda(1-F_{\varepsilon_c}(\varepsilon_{c,t}))\big)\mathbb{E}_t\left[\varepsilon_{c,t+1}^{-\theta}\frac{Y_{t+1}}{S_{t+1}/D_{t+1} + Y_{t+1}}\right]\notag\\
&\quad +\beta \mathbb{E}\left[\varepsilon_{c,t+1}^{-\theta}\big(1+ bF_{\varepsilon_c}(\varepsilon_{c,t+1})+ b \lambda(1-F_{\varepsilon_c}(\varepsilon_{c,t+1}))\big)\right]\bigg)R_{f,t+1} = \big(1+ bF_{\varepsilon_c}(\varepsilon_{c,t})+ b \lambda(1-F_{\varepsilon_c}(\varepsilon_{c,t}))\big).\label{eq:GM_Euler3}
\end{align}
Finally, defining
\begin{equation}
h(\varepsilon_{c,t},Y_t):= \frac{S_t}{D_t}\Big(1+ bF_{\varepsilon_c}(\varepsilon_{c,t})+ b \lambda(1-F_{\varepsilon_c}(\varepsilon_{c,t}))\Big)\left(1-\beta \mathbb{E}
\left[\varepsilon_{c,t}^{1-\theta} \right]\right), \label{eq:GM_hfun_Funof_PriceDivratio}
\end{equation}
recalling that
\begin{align*}
	R_{S,t+1} =\frac{S_{t+1}+D_{t+1}}{S_t} = \frac{S_{t+1}/D_{t+1}+1}{S_t/D_t}\cdot \frac{D_{t+1}}{D_t} = \frac{S_{t+1}/D_{t+1}+1}{S_t/D_t}\cdot \frac{Y_{t}}{Y_{t+1}}\varepsilon_{c,t+1},
\end{align*}
and plugging the above into \eqref{eq:GM_Euler1} and \eqref{eq:GM_Euler3},
we obtain the risk-free return \eqref{eq:GM_EquiRiskFree} and stock return \eqref{eq:GM_EquiStockReturn}: 
\begin{equation*}
\begin{split}
\hspace{-0.5cm} & R_{f,t+1}^{-1}\Big(1+ bF_{\varepsilon_c}(\varepsilon_{c,t})+ b \lambda(1-F_{\varepsilon_c}(\varepsilon_{c,t}))\Big) \\
\hspace{-0.5cm}=& \,\beta\mathbb{E}_t \left[  \varepsilon_{c,t+1}^{-\theta} \frac{Y_{t+1}}{Y_{t+1}+\frac{S_{t+1}}{D_{t+1}}} \right]\frac{\gamma \Big(F_{\varepsilon_c}(\varepsilon_{c,t}) + \lambda(1-F_{\varepsilon_c}(\varepsilon_{c,t}))\Big)}{1-\beta \mathbb{E} \left[\varepsilon_{c,t+1}^{1-\theta} \right]}\\
\hspace{-0.5cm}&+ \beta\mathbb{E}\left[  \varepsilon_{c,t+1}^{-\theta}   \Big( 1+ bF_{\varepsilon_c}(\varepsilon_{c,t+1})+ b \lambda(1-F_{\varepsilon_c}(\varepsilon_{c,t+1})) \Big) \right]\\
\hspace{-0.5cm}=& \, \beta\mathbb{E}_t \left[  \varepsilon_{c,t+1}^{-\theta} \frac{Y_{t+1}\Big( 1+ bF_{\varepsilon_c}(\varepsilon_{c,t+1})+ b \lambda(1-F_{\varepsilon_c}(\varepsilon_{c,t+1}))\Big)\left(1-\beta \mathbb{E}
	\left[\varepsilon_{c,t+1}^{1-\theta} \right]\right)}{Y_{t+1}\Big( 1+ bF_{\varepsilon_c}(\varepsilon_{c,t+1})+ b \lambda(1-F_{\varepsilon_c}(\varepsilon_{c,t+1}))\Big)\left(1-\beta \mathbb{E}
	\left[\varepsilon_{c,t+1}^{1-\theta} \right]\right)+h\left(\varepsilon_{c,t+1},Y_{t+1}\right)} \right]\\
\hspace{-0.5cm}&\times \frac{\gamma \Big(F_{\varepsilon_c}(\varepsilon_{c,t}) + \lambda(1-F_{\varepsilon_c}(\varepsilon_{c,t}))\Big)}{1-\beta \mathbb{E} \left[\varepsilon_{c,t+1}^{1-\theta} \right]}+ \beta\mathbb{E}_t \left[  \varepsilon_{c,t+1}^{-\theta}   \Big(  1+ bF_{\varepsilon_c}(\varepsilon_{c,t+1})+ b \lambda(1-F_{\varepsilon_c}(\varepsilon_{c,t+1}))\Big) \right];\\
\end{split}
\end{equation*}

\begin{equation*}
\begin{split}
R_{S,t+1} =& \left(\dfrac{h\left(\varepsilon_{c,t+1},Y_{t+1}\right)}{\Big( 1+ bF_{\varepsilon_c}(\varepsilon_{c,t+1})+ b \lambda(1-F_{\varepsilon_c}(\varepsilon_{c,t+1}))\Big)\left(1-\beta \mathbb{E}
	\left[\varepsilon_{c,t+1}^{1-\theta} \right]\right)}+1\right)\\
& \times \left(\frac{h\left(\varepsilon_{c,t},Y_{t}\right)}{\Big( 1+ bF_{\varepsilon_c}(\varepsilon_{c,t})+ b \lambda(1-F_{\varepsilon_c}(\varepsilon_{c,t}))\Big)\left(1-\beta \mathbb{E}
	\left[\varepsilon_{c,t+1}^{1-\theta} \right]\right)}\right)^{-1}\varepsilon_{d,t+1} \\[0.4cm]
=& \dfrac{\Big(1+ bF_{\varepsilon_c}(\varepsilon_{c,t})+ b \lambda(1-F_{\varepsilon_c}(\varepsilon_{c,t}))\Big)}{h\left(\varepsilon_{c,t},Y_{t}\right)}\\
 & \times \left( \dfrac{h\left(\varepsilon_{c,t+1},Y_{t+1}\right)}{\Big(1+ bF_{\varepsilon_c}(\varepsilon_{c,t+1})+ b \lambda(1-F_{\varepsilon_c}(\varepsilon_{c,t+1}))\Big)} + 1-\beta \mathbb{E}
	\left[\varepsilon_{c,t+1}^{1-\theta} \right] \right)\varepsilon_{d,t+1}.
\end{split}
\end{equation*}
\qed

\subsection{Consumption-wealth ratio in Model I} \label{appendix:CWratio_generalmodel}  

Recall the first order condition in \eqref{eq:BM_MarginalU}:
\begin{equation*}
U_t^{'}( \bar{C}_t) = \beta \mathbb{E}_t \left[ U_{t+1}^{'}( \bar{C}_{t+1})R_{S,t+1} \right].
\end{equation*}
In equilibrium, we have $W_{t} = S_{t} + \bar{C}_{t}$ and
\begin{equation*}
U'_t( \bar{C}_t) = m'( \bar{C}_t)\Big(1 + bF_{\varepsilon_c}(\varepsilon_{c,t}) + b\lambda\left(1-F_{\varepsilon_c}(\varepsilon_{c,t})\right)\Big) ,
\end{equation*}
thus we can write
\begin{equation*}
\begin{split}
\bar{C}_{t}^{-\theta} \Big(1 + bF_{\varepsilon_c}(\varepsilon_{c,t}) + b\lambda\left(1-F_{\varepsilon_c}(\varepsilon_{c,t})\right)\Big) = &\beta \mathbb{E}_t \left[\bar{C}_{t+1}^{-\theta} \Big(1 + bF_{\varepsilon_c}(\varepsilon_{c,t+1}) + b\lambda\left(1-F_{\varepsilon_c}(\varepsilon_{c,t+1})\right)\Big)R_{S,t+1} \right].
\end{split}
\end{equation*}
Now divide both sides by $\bar{C}_{t}^{-\theta}\Big(1 + bF_{\varepsilon_c}(\varepsilon_{c,t}) + b\lambda\left(1-F_{\varepsilon_c}(\varepsilon_{c,t})\right)\Big)$ and multiply by $\dfrac{S_{t}}{W_{t}} = \dfrac{W_{t} - \bar{C}_{t}}{W_t}$:
\begin{equation*}
 \dfrac{W_{t} - \bar{C}_{t}}{W_t} = \beta \mathbb{E}_t \left[ \dfrac{1 + bF_{\varepsilon_c}(\varepsilon_{c,t+1}) + b\lambda\left(1-F_{\varepsilon_c}(\varepsilon_{c,t+1})\right)}{1 + bF_{\varepsilon_c}(\varepsilon_{c,t}) + b\lambda\left(1-F_{\varepsilon_c}(\varepsilon_{c,t})\right)}\left(\dfrac{\bar{C}_{t+1}}{\bar{C}_t}\right)^{-\theta}\dfrac{W_{t+1}}{W_{t}} \right] .
\end{equation*}
Finally, given that in equilibrium $W_{t+1} = S_{t+1} + \bar{C}_{t+1}$ and that
\begin{equation}
\dfrac{D_{t+1}}{W_{t}} = \dfrac{D_{t+1}}{W_{t}} \cdot \dfrac{\bar{C}_{t+1}}{\bar{C}_{t+1}} \cdot \dfrac{\bar{C}_{t}}{\bar{C}_{t}} = \varepsilon_{c,t+1} \cdot \dfrac{D_{t+1}}{\bar{C}_{t+1}} \cdot \dfrac{\bar{C}_{t}}{W_{t}},
\end{equation}
by replacing the soluton for the price-dividend ratio $\frac{S_{t+1}}{D_{t+1}}$ and further reorganizing the terms, we derive the equilibrium consumption-wealth ratio:
\begin{equation*}
\begin{split}
\dfrac{\bar{C}_{t}}{W_t}   = & \left( \rule{0cm}{1cm} 1 +\beta \mathbb{E}_t \left[ \rule{0cm}{1cm}\varepsilon_{c,t+1}^{1-\theta} \dfrac{\big( 1+ bF_{\varepsilon_c}(\varepsilon_{c,t+1})+ b \lambda(1-F_{\varepsilon_c}(\varepsilon_{c,t+1})) \big)}{\big( 1+ bF_{\varepsilon_c}(\varepsilon_{c,t})+ b \lambda(1-F_{\varepsilon_c}(\varepsilon_{c,t
})) \big)} \right] \right. \\
& \left. \hspace{2cm} + \beta \mathbb{E}_t \left[  \rule{0cm}{1cm}  \varepsilon_{c,t+1}^{1-\theta}  \dfrac{ h(\varepsilon_{c,t+2}, Y_{t+2})}{Y_{t+1}\big( 1+ bF_{\varepsilon_c}(\varepsilon_{c,t})+ b \lambda(1-F_{\varepsilon_c}(\varepsilon_{c,t})) \big) \Big(1-\beta \mathbb{E} \left[\varepsilon_{c,t+2}^{-\theta}\right]\Big)} \right]\right)^{-1}. \\
\end{split}
\end{equation*}

\subsection{Proof of Corollary \ref{SDF_general_cor}}

From the proof of Theorem \ref{GM_thm}, we immediately derive the stochastic discount factor as the intertemporal marginal rate of substitution:
\begin{equation*}
\begin{split}
M_{t+1} & = \dfrac{\beta U'_{t+1}( \bar{C}_{t+1})}{U'_t( \bar{C}_t)} \\
& = \frac{\beta}{1+ bF_{\varepsilon_c}(\varepsilon_{c,t})+ b \lambda(1-F_{\varepsilon_c}(\varepsilon_{c,t})) } \Bigg ( \varepsilon_{c,t+1}^{-\theta} \big( 1+ bF_{\varepsilon_c}(\varepsilon_{c,t+1})+ b \lambda(1-F_{\varepsilon_c}(\varepsilon_{c,t+1})) \big) \Bigg. \\[0.2cm]
& \quad \Bigg. +  \gamma\beta    \varepsilon_{c,t+1}^{-\theta}Y_{t+1}\dfrac{\big(F_{\varepsilon_c}(\varepsilon_{c,t}) + \lambda(1-F_{\varepsilon_c}(\varepsilon_{c,t})) \big)\big(1+ bF_{\varepsilon_c}(\varepsilon_{c,t+1})+ b \lambda (1-F_{\varepsilon_c}(\varepsilon_{c,t+1}))\big)}{Y_{t+1}\Big(1+ bF_{\varepsilon_c}(\varepsilon_{c,t+1})+  b \lambda(1-F_{\varepsilon_c}(\varepsilon_{c,t+1}))\Big)\left(1-\beta \mathbb{E}
\left[\varepsilon_{c,t+1}^{1-\theta} \right]\right)+h(\varepsilon_{c,t+1},Y_{t+1})} \Bigg).
\end{split}
\end{equation*}

\subsection{Proof of Proposition \ref{prop:FixedPoint}}

Suppose in equilibrium that the price-dividend ratio is
\begin{equation*}
\frac{S_{t+1}}{D_{t+1}}=f\left(\varepsilon_{c,t+1},Y_{t+1}\right)= \dfrac{\tilde{f}\left(\varepsilon_{c,t+1},Y_{t+1}\right)}{1+ bF_{\varepsilon_c}(\varepsilon_{c,t+1})+ b \lambda(1-F_{\varepsilon_c}(\varepsilon_{c,t+1}))}.\\[0.1cm]
\end{equation*}
Additionally, assume that the consumption growth rates $\varepsilon_{c,t}, t=0,1,\dots$ are i.i.d, independent of $Y_{t}$ and satisfy the following growth condition:
\begin{equation*}
\beta \mathbb{E} \left[ \varepsilon_{c,t+1}^{1-\theta} \right] < 1.
\end{equation*}
Given these assumptions, from the Euler equation in \eqref{eq:GM_Euler1} we can compute
\begin{equation*}
\begin{split}
\tilde{f}\left(\varepsilon_{c,t},Y_{t}\right) = & \, \beta\mathbb{E}_t
\left[ \rule{0cm}{1cm}  \varepsilon_{c,t+1}^{1-\theta}
\frac{1+ bF_{\varepsilon_c}(\varepsilon_{c,t+1})+ b \lambda(1-F_{\varepsilon_c}(\varepsilon_{c,t+1}))+\tilde{f}\left(\varepsilon_{c,t+1},Y_{t+1}\right)}{Y_{t+1}\Big(1+ bF_{\varepsilon_c}(\varepsilon_{c,t+1})+ b \lambda(1-F_{\varepsilon_c}(\varepsilon_{c,t+1}))\Big)+\tilde{f}\left(\varepsilon_{c,t+1},Y_{t+1}\right)}  \right]  \\
& \times \rule{0cm}{1cm} \frac{\gamma \Big(F_{\varepsilon_c}(\varepsilon_{c,t}) + \lambda(1-F_{\varepsilon_c}(\varepsilon_{c,t}))  \Big)}{1-\beta \mathbb{E}
\left[\varepsilon_{c,t+1}^{1-\theta} \right]} Y_{t}\\
& + \beta \mathbb{E}_t
\left[  \varepsilon_{c,t+1}^{1-\theta} \dfrac{Y_{t}}{Y_{t+1}}
\left( \tilde{f}\left(\varepsilon_{c,t+1},Y_{t+1}\right) +1+ bF_{\varepsilon_c}(\varepsilon_{c,t+1})+ b \lambda(1-F_{\varepsilon_c}(\varepsilon_{c,t+1}))\right) \right].
\end{split}
\end{equation*}
We now show that $h(\varepsilon_{c,t},Y_t) := \tilde{f}\left(\varepsilon_{c,t},Y_{t}\right)\left(1-\beta \mathbb{E}
\left[\varepsilon_{c,t+1}^{1-\theta} \right]\right)$ is the unique fixed point of the following equation:
\begin{equation*}
\begin{split}
&h(\varepsilon_{c,t},Y_t) = \gamma\Big(F_{\varepsilon_c}(\varepsilon_{c,t}) + \lambda(1-F_{\varepsilon_c}(\varepsilon_{c,t}))  \Big)Y_t \\
& \times \beta\mathbb{E}_t
\left[  \varepsilon_{c,t+1}^{1-\theta}\frac{\Big(1+ bF_{\varepsilon_c}(\varepsilon_{c,t+1})+ b \lambda(1-F_{\varepsilon_c}(\varepsilon_{c,t+1}))\Big)\left(1-\beta \mathbb{E}
\left[\varepsilon_{c,t+1}^{1-\theta} \right]\right)+h(\varepsilon_{c,t+1},Y_{t+1})}{Y_{t+1}\Big(1+ bF_{\varepsilon_c}(\varepsilon_{c,t+1})+ b \lambda(1-F_{\varepsilon_c}(\varepsilon_{c,t+1}))\Big)\left(1-\beta \mathbb{E}
\left[\varepsilon_{c,t+1}^{1-\theta} \right]\right)+h(\varepsilon_{c,t+1},Y_{t+1})}  \right]  \\
& +  \beta\mathbb{E}_t
\left[  \varepsilon_{c,t+1}^{1-\theta}\frac{Y_{t}}{Y_{t+1}}
\Bigg( h(\varepsilon_{c,t+1},Y_{t+1})+ \Big(1+ bF_{\varepsilon_c}(\varepsilon_{c,t+1})+ b \lambda(1-F_{\varepsilon_c}(\varepsilon_{c,t+1}))\Big) \bigg(1-\beta \mathbb{E}
\left[\varepsilon_{c,t+1}^{1-\theta} \right]\bigg)\Bigg) \right]. \\[0.2cm]
\end{split}
\end{equation*}
Since the normed vector space of continuous functions is a complete metric space, we hereby prove that $h$ is a contraction mapping. To this end, we define the mapping
\begin{equation*}
\begin{split}
 & \mathbb{T}h(x,y) := \;   \gamma\Big(F_{\varepsilon_c}(x) + \lambda(1-F_{\varepsilon_c}(x)) \Big)y \\
 & \times  \beta\mathbb{E}_t
\left[ \varepsilon_{c,t+1}^{1-\theta}\frac{\Big(1+ bF_{\varepsilon_c}(\varepsilon_{c,t+1})+ b \lambda(1-F_{\varepsilon_c}(\varepsilon_{c,t+1}))\Big)\left(1-\beta \mathbb{E}
\left[\varepsilon_{c,t+1}^{1-\theta} \right]\right)+h(\varepsilon_{c,t+1},Y_{t+1})}{Y_{t+1}\Big(1+ bF_{\varepsilon_c}(\varepsilon_{c,t+1})+ b \lambda(1-F_{\varepsilon_c}(\varepsilon_{c,t+1}))\Big)\left(1-\beta \mathbb{E}
\left[\varepsilon_{c,t+1}^{1-\theta} \right]\right)+h(\varepsilon_{c,t+1},Y_{t+1})}  \right]  \\[0.2cm]
 & +  \beta\mathbb{E}_t
\left[  \varepsilon_{c,t+1}^{1-\theta}\frac{y}{Y_{t+1}}
\Bigg( h(\varepsilon_{c,t+1},Y_{t+1})+ \Big(1+ bF_{\varepsilon_c}(\varepsilon_{c,t+1})+ b \lambda(1-F_{\varepsilon_c}(\varepsilon_{c,t+1}))\Big) \bigg(1-\beta \mathbb{E}
\left[\varepsilon_{c,t+1}^{1-\theta} \right]\bigg)\Bigg)  \right],\\[0.2cm]
\end{split}
\end{equation*}
and compute its first order derivative:
\begin{equation*}
\begin{split}
\frac{\partial \mathbb{T}h(x,y)}{\partial h} = \; & \beta\mathbb{E}_t
\left[  \varepsilon_{c,t+1}^{1-\theta}
\frac{y}{Y_{t+1}} \right] + \gamma\Big(F_{\varepsilon_c}(x) + \lambda(1-F_{\varepsilon_c}(x)) \Big)y 
\\
& \hspace{-1cm} \times \beta \mathbb{E}_t
\left[ \varepsilon_{c,t+1}^{1-\theta}\frac{\Big(1+ bF_{\varepsilon_c}(\varepsilon_{c,t+1})+ b \lambda(1-F_{\varepsilon_c}(\varepsilon_{c,t+1}))\Big)\left(1-\beta \mathbb{E}
\left[\varepsilon_{c,t+1}^{1-\theta} \right]\right)\left(Y_{t+1}-1\right)}{\left( Y_{t+1}\Big(1+ bF_{\varepsilon_c}(\varepsilon_{c,t+1})+ b \lambda(1-F_{\varepsilon_c}(\varepsilon_{c,t+1}))\Big)\left(1-\beta \mathbb{E}
\left[\varepsilon_{c,t+1}^{1-\theta} \right]\right)+h(\varepsilon_{c,t+1},Y_{t+1})\right)^2}  \right].  \\[0.2cm]
\end{split}
\end{equation*}
Because $Y_{t+1} \geq 1$, $\dfrac{\partial \mathbb{T}h(x,y)}{\partial h} $ is always positive, so the mapping is increasing in $h$, decreasing in $\varepsilon_{c,t}$, and the first order derivative is decreasing in $h$. 

While we could not prove that $\mathbb{T}h(x,y)$ is a contraction for any $h \geq 0$, we can show that it is a contraction when $h$ is greater than a lower bound $\underline{h}$ satisfying
\begin{equation*}
\mathbb{E}_{t}
\left[\left( \beta \varepsilon_{c,t+1}^{1-\theta}\dfrac{y}{Y_{t+1}}\frac{\gamma \lambda \Big(1+ bF_{\varepsilon_c}(\varepsilon_{c,t+1})+ b \lambda(1-F_{\varepsilon_c}(\varepsilon_{c,t+1}))\Big)\left(1-\beta \mathbb{E}
\left[\varepsilon_{c,t+1}^{1-\theta} \right]\right)}{\left(\Big(1+ bF_{\varepsilon_c}(\varepsilon_{c,t+1})+ b \lambda(1-F_{\varepsilon_c}(\varepsilon_{c,t+1}))\Big)\left(1-\beta \mathbb{E}
\left[\varepsilon_{c,t+1}^{1-\theta} \right]\right)+\underline{h}\right)^2} + \beta \varepsilon_{c,t+1}^{1-\theta} \dfrac{y}{Y_{t+1}}\right)^2 \right] = 1.\\[0.2cm]
\end{equation*}
Under this assumption, it thus follows that
\begin{equation*}
\begin{split}
\mathbb{T}h(x,y)  \geq \; & \mathbb{T}0\\
= \; & \gamma\Big(F_{\varepsilon_c}(x) + \lambda(1-F_{\varepsilon_c}(x))\Big) \beta \mathbb{E}_{t}
\left[  \varepsilon_{c,t+1}^{1-\theta} \dfrac{y}{Y_{t+1}} \right]  \\
& +  \beta\mathbb{E}_{t}\left[  \varepsilon_{c,t+1}^{1-\theta} \dfrac{y}{Y_{t+1}}
\Big(1+ bF_{\varepsilon_c}(\varepsilon_{c,t+1})+ b \lambda(1-F_{\varepsilon_c}(\varepsilon_{c,t+1})) \Big)\Big(1-\beta \mathbb{E}_{t}
\left[\varepsilon_{c,t+1}^{1-\theta} \right]\Big) \right] \\[0.1cm]
\geq \; & \gamma  \beta\mathbb{E}_{t}
\left[ \varepsilon_{c,t+1}^{1-\theta} \dfrac{y}{Y_{t+1}}\right] + \beta \mathbb{E}_{t}
\left[  \varepsilon_{c,t+1}^{1-\theta}\dfrac{y}{Y_{t+1}}
\Big(1+ bF_{\varepsilon_c}(\varepsilon_{c,t+1})+ b \lambda(1-F_{\varepsilon_c}(\varepsilon_{c,t+1})) \Big)\bigg(1-\beta \mathbb{E}
\left[\varepsilon_{c,t+1}^{1-\theta} \right]\bigg) \right] \\[0.1cm]
\geq \; & \gamma\beta \,\mathbb{E}_{t}
\left[  \varepsilon_{c,t+1}^{1-\theta} \dfrac{y}{Y_{t+1}} \right],
\end{split}
\end{equation*}
indicating that $\mathbb{T}(\mathbb{T}h(x,y)) \geq \mathbb{T}(\mathbb{T}0)$; starting from $0$, the iteration is increasing and the function has a lower-bound.

We can finally show that the mapping is a contraction mapping when $h>\underline{h}$:
\begin{equation*}
\begin{split}
& \Bigg| \mathbb{T}h_1(x,y) - \mathbb{T}h_2(x,y) \Bigg| \\
= \; & \Bigg| \beta\mathbb{E}_t
\left[ \varepsilon_{c,t+1} ^{1-\theta} \dfrac{y}{Y_{t+1}}\Big(h_1(\varepsilon_{c,t+1},Y_{t+1})- h_2(\varepsilon_{c,t+1},Y_{t+1})\Big) \bigg(1-\beta \mathbb{E}
\left[\varepsilon_{c,t+1}^{1-\theta} \right]\bigg)\right] \\
& +  \gamma\Big(F_{\varepsilon_c}(x) + \lambda(1-F_{\varepsilon_c}(x)) \Big) y   \\
& \times \beta \mathbb{E}_{t}
\Bigg[\varepsilon_{c,t+1} ^{1-\theta}\frac{ \Big(1+ bF_{\varepsilon_c}(\varepsilon_{c,t+1})+ b \lambda(1-F_{\varepsilon_c}(\varepsilon_{c,t+1}))\Big)\left(Y_{t+1}-1 \right)\left(1-\beta \mathbb{E}
\left[\varepsilon_{c,t+1}^{1-\theta} \right]\right)}{ Y_{t+1}\Big(1+ bF_{\varepsilon_c}(\varepsilon_{c,t+1})+ b \lambda(1-F_{\varepsilon_c}(\varepsilon_{c,t+1}))\Big)\left(1-\beta \mathbb{E}
\left[\varepsilon_{c,t+1}^{1-\theta} \right]\right)+h_1(\varepsilon_{c,t+1},Y_{t+1})} \Bigg.  \\
& \Bigg. \times \frac{h_1(\varepsilon_{c,t+1},Y_{t+1})- h_2(\varepsilon_{c,t+1},Y_{t+1})}{ Y_{t+1}\Big(1+ bF_{\varepsilon_c}(\varepsilon_{c,t+1})+ b \lambda(1-F_{\varepsilon_c}(\varepsilon_{c,t+1}))\Big)\left(1-\beta \mathbb{E}
\left[\varepsilon_{c,t+1}^{1-\theta} \right]\right)+h_2(\varepsilon_{c,t+1},Y_{t+1})}\Bigg]  \Bigg| \\[0.2cm]
\leq \; & \Bigg|\beta \mathbb{E}_t
\Bigg[ \varepsilon_{c,t+1} ^{1-\theta} \dfrac{y}{Y_{t+1}}\Big(h_1(\varepsilon_{c,t+1},Y_{t+1})- h_2(\varepsilon_{c,t+1},Y_{t+1})\Big)  \bigg(1-\beta \mathbb{E}
\left[\varepsilon_{c,t+1}^{1-\theta} \right]\bigg) \Bigg]\\
& +  \gamma \lambda \beta\, \mathbb{E}_t
\Bigg[ \varepsilon_{c,t+1} ^{1-\theta}  y\Big(h_1(\varepsilon_{c,t+1},Y_{t+1})- h_2(\varepsilon_{c,t+1},Y_{t+1})\Big) \\
&  \times \frac{\Big(1+ bF_{\varepsilon_c}(\varepsilon_{c,t+1})+ b \lambda(1-F_{\varepsilon_c}(\varepsilon_{c,t+1}))\Big)\left(Y_{t+1}-1 \right)\left(1-\beta \mathbb{E}
\left[\varepsilon_{c,t+1}^{1-\theta} \right]\right)}{ \bigg(Y_{t+1}\Big(1+ bF_{\varepsilon_c}(\varepsilon_{c,t+1})+ b \lambda(1-F_{\varepsilon_c}(\varepsilon_{c,t+1}))\Big)\left(1-\beta \mathbb{E}
\left[\varepsilon_{c,t+1}^{1-\theta} \right]\right)+\underline{h}\bigg)^2 }\Bigg]  \Bigg| \\[0.2cm]
\leq \; & \beta\mathbb{E}_t
\Bigg[\Bigg(  \varepsilon_{c,t+1}^{1-\theta}\dfrac{y}{Y_{t+1}} \frac{\gamma \lambda \Big(1+ bF_{\varepsilon_c}(\varepsilon_{c,t+1})+ b \lambda(1-F_{\varepsilon_c}(\varepsilon_{c,t+1}))\Big)\left(1-\beta \mathbb{E}
\left[\varepsilon_{c,t+1}^{1-\theta} \right]\right)}{\left(\Big(1+ bF_{\varepsilon_c}(\varepsilon_{c,t+1})+ b \lambda(1-F_{\varepsilon_c}(\varepsilon_{c,t+1}))\Big)\left(1-\beta \mathbb{E}
\left[\varepsilon_{c,t+1}^{1-\theta} \right]\right)+\underline{h}\right)^2}+ \varepsilon_{c,t+1}^{1-\theta} \dfrac{y}{Y_{t+1}}\Bigg)^2 \Bigg]^{1/2} \\
&\times \mathbb{E}_t \left[\Big(h_1(\varepsilon_{c,t+1},Y_{t+1})- h_2(\varepsilon_{c,t+1},Y_{t+1})\Big)^2 \right]^{1/2}\\[0.2cm]
\leq \; & \mathbb{E}_t \left[\Big(h_1(\varepsilon_{c,t+1},Y_{t+1})- h_2(\varepsilon_{c,t+1},Y_{t+1})\Big)^2 \right]^{1/2}.
\end{split}
\end{equation*}
This yields that
\begin{equation*}
\begin{split}
& \mathbb{E}_t \left[\left( \mathbb{T}h_1(\varepsilon_{c,t},Y_{t}) - \mathbb{T}h_2(\varepsilon_{c,t},Y_{t}) \right)^2 \right]^{1/2} \leq \mathbb{E}_t \left[ \left(h_1(\varepsilon_{c,t+1},Y_{t+1})- h_2(\varepsilon_{c,t+1},Y_{t+1})\right)^2  \right]^{1/2},
\end{split}
\end{equation*}
so $\mathbb{T}h(x,y)$ has a unique fixed point for $h>\underline{h}$. \qed

\subsection{Proof of Theorem \ref{Baseline_thm}} \label{subsec:ProofBaseline_thm}
The proof is similar to that of Theorem \ref{GM_thm}. For completeness and readability, we provide a self-contained version here. 

In the absence of prospective gain-loss utility, we can consider Markovian strategies with Markovian state variables $(W_t, \bar{C}_t, D_t, \varepsilon_{c,t})$. Again, we denote by $f(W_t, \bar{C}_t, D_t, \varepsilon_{c,t})$  the optimal value of the agent's portfolio selection problem. 

By the dynamic programming principle, the optimization problem can be written as follows:
\begin{eqnarray}\label{eq:BM_DP}
f(W_t, \bar{C}_t, D_t, \varepsilon_{c,t})= \underset{C_t,\alpha_t}{\max}\ \ U_t(C_t) +\beta \mathbb{E}_t \left[ f(W_{t+1}, \bar{C}_{t+1}, D_{t+1}, \varepsilon_{c,t+1}) \right],
\end{eqnarray}	
subject to the wealth equation
\begin{equation*}
W_{t+1} = (W_t-C_t) \left( \alpha_t R_{S,t+1}+(1-\alpha_t)R_{f,t+1} \right) +L_{t+1},
\end{equation*}
and the labor income dynamics $L_{t+1} =  \bar{C}_t \varepsilon_{c,t+1} - D_t\varepsilon_{d,t+1}$.

Denote by $C^\star_t$ and $\alpha^\star_t$ the optimal consumption and allocation in the stock, and recall that in  equilibrium $C^\star_t =  \bar{C}_t$ and $\alpha^\star_t = 1$. Then, the first order conditions for optimality can be written as follows:
\begin{eqnarray}\label{BM_FONC}
\begin{aligned}
\begin{cases}
U_t^{'}(C_t^\star) - \beta \mathbb{E}_t \left[ f_{W_t}(W_{t+1}^\star, \bar{C}_{t+1}, D_{t+1}, \varepsilon_{c,t+1})R_{S,t+1} \right] = 0,
\\
\beta \mathbb{E}_t \left[ f_{W_t}(W_{t+1}^\star, \bar{C}_{t+1}, D_{t+1}, \varepsilon_{c,t+1})(W_t-\bar{C}_t)(R_{S,t+1}-R_{f,t+1}) \right] = 0,
\end{cases}
\end{aligned}
\end{eqnarray}
where $W_{t+1}^\star$ stands for the wealth associated with the optimal strategy. Moreover, we have
\begin{equation*}
\begin{split}
& f(W_t, \bar{C}_t, D_t, \varepsilon_{c,t}) 	=  U_t(C_t^{\star}) +\beta \mathbb{E}_t \left[ f(W^{\star}_{t+1},\bar{C}_{t+1}, D_{t+1},\varepsilon_{c,t+1}) \right].\\
\end{split}
\end{equation*}
Taking the derivative with respect to $W_t$ on both sides of the above equation and recalling the wealth equation, we derive
\begin{equation*}
\begin{split}
 f_{W_t}(W_t, \bar{C}_t, D_t, \varepsilon_{c,t}) & =  U^{'}(C_t^{\star})\frac{\partial C^\star}{\partial W_t}(W_t,\bar{C}_{t},D_{t},\varepsilon_{c,t}) \\
& \quad + \beta \mathbb{E}_t \biggl[ f_{W_t}(W^{\star}_{t+1},\bar{C}_{t+1}, D_{t+1},\varepsilon_{c,t+1}) \left( 1-\frac{\partial C^\star}{\partial W_t}(W_t,\bar{C}_{t},D_{t},\varepsilon_{c,t}) \right) R_{S,t+1} \biggr]\\[5pt]
&= U_t^{'}(C_t^{\star}),
\end{split}
\end{equation*}
where the second equality follows from the first equation of \eqref{BM_FONC}. Plugging the above into \eqref{BM_FONC}, recalling again that under market equilibrium $C_t^\star =  \bar{C}_t$, and observing that $W_t-\bar{C}_t$ is known at time $t$, we derive the following Euler equations:
\begin{eqnarray}\label{BM_EulerG}
\begin{aligned}
\begin{cases}
U_t^{'}( \bar{C}_t) - \beta \mathbb{E}_t \left[ U_{t+1}^{'}( \bar{C}_{t+1})R_{S,t+1} \right] = 0,
\\
\beta \mathbb{E}_t \left[ U_{t+1}^{'}( \bar{C}_{t+1})(R_{S,t+1}-R_{f,t+1}) \right] = 0.
\end{cases}
\end{aligned}
\end{eqnarray}
Next, from \eqref{baseline_utility}, we have
\begin{align}\label{eq:BM_MarginalU}
	U'_t(C_t) = \left(1 + bF_{t-1,\tilde{C}_{t-1,t}}(C_t) + b\lambda\left(1-F_{t-1,\tilde{C}_{t-1,t}}(C_t)\right)\right)m'(C_t),
\end{align}
where $F_{t-1,\tilde{C}_{t-1,t}}$ denotes the CDF of $\tilde{C}_{t-1,t}$ conditional on ${\cal F}_{t-1}$. In addition, since $ \bar{C}_t =  \bar{C}_{t-1}\varepsilon_{c,t}$, we derive from \eqref{eq:BM_MarginalU} that
\begin{align*}
U'_t( \bar{C}_t) = \Big(1 + bF_{\varepsilon_c}(\varepsilon_{c,t}) + b\lambda\left(1-F_{\varepsilon_c}(\varepsilon_{c,t})\right)\Big)m'( \bar{C}_t).
\end{align*}
Finally, recalling the form of the utility function $m$ in \eqref{eq:power_utility_function}, plugging the above expression for $U'_t( \bar{C}_t)$ into \eqref{BM_EulerG}, and using the growth condition $\beta \mathbb{E}_t \left[\varepsilon_{c,t+1}^{-\theta}\varepsilon_{d,t+1} \right] < 1$, we obtain by standard calculations the risk-free return \eqref{eq:RiskFreeReturn_BaselineModel}, the stock price-dividend ratio \eqref{eq:StockPriceDividendRatio_BaselineModel}, the stock return \eqref{eq:StockReturn:BaselineModel}, and the risk premium \eqref{eq:RiskPremium_BaselineModel}.
 \qed

\subsection{Consumption-wealth ratio in Model II} \label{appendix:CWratio_baselinemodel}

Following the same lines of argument as in Appendix \ref{appendix:CWratio_generalmodel}, the consumption-dividend ratio for Model II is given by
\begin{equation*}
\begin{split}
\dfrac{\bar{C}_{t}}{W_t}   = & \left( \rule{0cm}{1cm} 1 +\beta \mathbb{E}_t \left[ \rule{0cm}{1cm}\left(\dfrac{\bar{C}_{t+1}}{\bar{C}_{t}}\right)^{1-\theta} \dfrac{\big( 1+ bF_{\varepsilon_c}(\varepsilon_{c,t+1})+ b \lambda(1-F_{\varepsilon_c}(\varepsilon_{c,t+1})) \big)}{\big( 1+ bF_{\varepsilon_c}(\varepsilon_{c,t})+ b \lambda(1-F_{\varepsilon_c}(\varepsilon_{c,t
})) \big)} \right] \right. \\
& \left.  \quad + \beta  \mathbb{E}_t \left[ \rule{0cm}{1cm} \bar{C}_{t+1}^{-\theta} D_{t+1}   \dfrac{ \beta  \mathbb{E}_{t+1} \bigg[\varepsilon_{c,t+2}^{-\theta}\varepsilon_{d,t+2}\big( 1+ bF_{\varepsilon_c}(\varepsilon_{c,t+2})+ b \lambda(1-F_{\varepsilon_c}(\varepsilon_{c,t+2}))\big)\bigg]}{\bar{C}_{t}^{1-\theta}\big( 1+ bF_{\varepsilon_c}(\varepsilon_{c,t})+ b \lambda(1-F_{\varepsilon_c}(\varepsilon_{c,t})) \big) \Big(1-\beta \mathbb{E}_{t+1} \left[\varepsilon_{c,t+2}^{-\theta}\varepsilon_{d,t+2}\right]\Big)} \right]\right)^{-1}. 
\end{split}
\end{equation*}
Compared to Model I, the difference lies in the second expectation, which here does not account for prospective gain-loss utilities through the function $h(\cdot, \cdot)$ defined in \eqref{eq:GM_hfun_Funof_PriceDivratio} - and more explicitly in \eqref{eq:GM_hFun}.

\subsection{Proof of Corollary \ref{SDF_baseline_cor}}

The proof is the same as of Corollary \ref{SDF_general_cor}.\qed

\subsection{Proof of Proposition \ref{prop:BaselineModel_ComparativeConsGrowth}}

From \eqref{eq:RiskFreeReturn_BaselineModel} and \eqref{eq:StockPriceDividendRatio_BaselineModel}, respectively, we derive that
\begin{equation*}
\begin{split}
& \frac{\partial R_{f,t+1} }{\partial \varepsilon_{c,t}} =
\frac{ b(1 - \lambda)F_{\varepsilon_{c}}'(\varepsilon_{c,t})  }{\beta\mathbb{E} \left[  \varepsilon_{c,t+1}^{-\theta}   \Big( 1+ bF_{\varepsilon_c}(\varepsilon_{c,t+1})+ b \lambda(1-F_{\varepsilon_c}(\varepsilon_{c,t+1}))\Big) \right]} < 0, \\[0.2cm]
\end{split}
\end{equation*}
and
\begin{equation*}
\begin{split}
& \frac{\partial \dfrac{S_t}{D_t} }{\partial \varepsilon_{c,t}} = -\frac{ \beta\mathbb{E}_{t} \bigg[\varepsilon_{c,t+1}^{-\theta}\varepsilon_{d,t+1}\Big(1+ bF_{\varepsilon_c}(\varepsilon_{c,t+1})+ b \lambda(1-F_{\varepsilon_c}(\varepsilon_{c,t+1}))\Big)\bigg]b(1 - \lambda)F_{\varepsilon_{c}}'(\varepsilon_{c,t}) }{\Big(1+ bF_{\varepsilon_c}(\varepsilon_{c,t})+ b \lambda(1-F_{\varepsilon_c}(\varepsilon_{c,t}))\Big)^2 \Big(1-\beta \mathbb{E}_{t}\Big[\varepsilon_{c,t+1}^{-\theta}\varepsilon_{d,t+1}\Big]\Big)} > 0.
\end{split}
\end{equation*}
Similarly, from 
\eqref{eq:StockReturn:BaselineModel}, we derive that
\begin{equation*}
\begin{split}
& \frac{\partial R_{S,t+1} }{\partial \varepsilon_{c,t}} \propto^{+} b(1 - \lambda)F_{\varepsilon_{c}}'(\varepsilon_{c,t}) \varepsilon_{d,t+1},
\end{split}
\end{equation*}
which is positive if $\varepsilon_{d,t+1} < 0$ and negative otherwise. (Here $\propto^{+}$ denotes proportionality with respect to some positive factor.)
\qed

\subsection{Proof of Proposition \ref{prop:BaselineModel_ComparativeParametersGainLoss}}
From \eqref{eq:RiskFreeReturn_BaselineModel}, we obtain that
\begin{equation*}
\begin{split}
& \frac{\partial R_{f,t+1} }{\partial b} =
\frac{ \beta(1 - \lambda)  \left( \mathbb{E} \Big[ \varepsilon_{c,t+1}^{-\theta}  F_{\varepsilon_c}(\varepsilon_{c,t+1}) \Big]  -  F_{\varepsilon_c}(\varepsilon_{c,t})\mathbb{E} \Big[ \varepsilon_{c,t+1}^{-\theta}  \Big]   \right) }{\bigg(\beta\mathbb{E} \left[  \varepsilon_{c,t+1}^{-\theta}   \Big( 1+ bF_{\varepsilon_c}(\varepsilon_{c,t+1})+ b \lambda(1-F_{\varepsilon_c}(\varepsilon_{c,t+1}))\Big) \right]\bigg)^2}, \\[0.2cm]
& \frac{\partial R_{f,t+1}}{\partial \lambda} = \frac{\beta b(1 +b)  \left( \mathbb{E} \Big[ \varepsilon_{c,t+1}^{-\theta} F_{\varepsilon_c}(\varepsilon_{c,t+1}) \Big]  -  F_{\varepsilon_c}(\varepsilon_{c,t})\mathbb{E} \Big[ \varepsilon_{c,t+1}^{-\theta} \Big]   \right)}{\bigg( \beta\mathbb{E} \left[ \varepsilon_{c,t+1}^{-\theta}   \Big( 1+ bF_{\varepsilon_c}(\varepsilon_{c,t+1})+ b \lambda(1-F_{\varepsilon_c}(\varepsilon_{c,t+1})) \Big) \right]\bigg)^2}.	
\end{split}
\end{equation*}
As a result, when $F_{\varepsilon_c}(\varepsilon_{c,t})  \geq \frac{\mathbb{E} \left[\varepsilon_{c,t+1}^{-\theta}F_{\varepsilon_c}(\varepsilon_{c,t+1}) \right]}{\mathbb{E}\left[\varepsilon_{c,t+1}^{-\theta}\right]}$, we have $\dfrac{\partial R_{f,t+1}}{\partial b}> 0$ and $	\dfrac{\partial R_{f,t+1}}{\partial \lambda}>0$; otherwise, $\dfrac{\partial R_{f,t+1}}{\partial b}< 0$ and $\dfrac{\partial R_{f,t+1}}{\partial \lambda}<0$.
Similarly, from \eqref{eq:StockPriceDividendRatio_BaselineModel}, we obtain that
\begin{equation*}
\begin{split}
& \frac{\partial \dfrac{S_t}{D_t} }{\partial b} = \frac{\beta(1 - \lambda)  \left( \mathbb{E}_{t}\Big[ \varepsilon_{c,t+1}^{-\theta} \varepsilon_{d,t+1} F_{\varepsilon_c}(\varepsilon_{c,t+1}) \Big]  -  F_{\varepsilon_c}(\varepsilon_{c,t})\mathbb{E}_{t} \Big[ \varepsilon_{c,t+1}^{-\theta}\varepsilon_{d,t+1}  \Big]   \right)}{\Big(1+ bF_{\varepsilon_c}(\varepsilon_{c,t})+ b \lambda(1-F_{\varepsilon_c}(\varepsilon_{c,t}))\Big)^2 \Big(1-\beta \mathbb{E}_{t}\Big[\varepsilon_{c,t+1}^{-\theta}\varepsilon_{d,t+1}\Big]\Big)},\\
& \frac{\partial \dfrac{S_t}{D_t} }{\partial \lambda} = \frac{\beta b(1 +b)  \left( \mathbb{E}_{t} \Big[ \varepsilon_{c,t+1}^{-\theta} \varepsilon_{d,t+1} F_{\varepsilon_c}(\varepsilon_{c,t+1}) \Big]  -  F_{\varepsilon_c}(\varepsilon_{c,t})\mathbb{E}_{t} \Big[ \varepsilon_{c,t+1}^{-\theta}\varepsilon_{d,t+1}  \Big]   \right)}{\Big(1+ bF_{\varepsilon_c}(\varepsilon_{c,t})+ b \lambda(1-F_{\varepsilon_c}(\varepsilon_{c,t}))\Big)^2 \left(1-\beta \mathbb{E}_{t}\Big[\varepsilon_{c,t+1}^{-\theta}\varepsilon_{d,t+1}\Big]\right)}.
\end{split}
\end{equation*}
In this case, when $F_{\varepsilon_c}(\varepsilon_{c,t})  \geq \frac{\mathbb{E}_{t} \Big[\varepsilon_{c,t+1}^{-\theta}\varepsilon_{d,t+1}F_{\varepsilon_c}(\varepsilon_{c,t+1}) \Big]}{\mathbb{E}_{t}\Big[\varepsilon_{c,t+1}^{-\theta}\varepsilon_{d,t+1}\Big]}$, we have $\dfrac{\partial \frac{S_t}{D_t}}{\partial b} > 0$ and $\dfrac{\partial \frac{S_t}{D_t} }{\partial \lambda}>0$; otherwise, we have $\dfrac{\partial \frac{S_t}{D_t} }{\partial b} < 0$ and $\dfrac{\partial \frac{S_t}{D_t} }{\partial \lambda}<0$. \qed

\section{Further numerical results} \label{appendix:AdditionalNumerics}

For reference, additional experiments for Model I are provided in the table below. In particular, compared to Section \ref{sec:Numerics}, we set a different level of risk aversion $\theta = 2$.   

\begin{table}[h!]
\centering
\begin{tabular}{lcccccc}
\hline \hline
\qquad & $\gamma=0$ & $\gamma=0.01$ & $\gamma=0.05$ & $\gamma = 0.1$  & $\gamma = 1$ & Empirical value\\
\hline
Risk-free rate &&&&&& \\
\quad Mean & 1.126 & 1.084  & 1.058  & 1.05  & 1.042 &  1.033 \\
\quad Standard deviation & 0.13 & 0.117  &  0.109 & 0.124  & 0.171 & 0.031  \\
Price-dividend ratio &&&&&& \\
\quad Mean & 13.671  & 25.427 & 97.04 &  193.511   & 1943.48 & 25.5   \\
\quad Standard deviation &  1.605  & 4.069 &  30.379 & 67.059   &  757.107 &   7.1   \\
Equity risk premium &&&&&& \\
\quad Mean & 0.049 &  0.048 & 0.037  & 0.035 &  0.024 & 0.085  \\
\quad Standard deviation & 0.259 &  0.224 & 0.191 & 0.192  & 0.138 &  0.201 \\
\hline \hline
\end{tabular}
\caption{Equilibrium prices in Model I as function of the prospective gain-loss utility weight $\gamma$. Other parameters as set as follows: $\theta = 2,b = 1, \lambda = 2, \beta = 0.98$. Empirical values are based on annualized data from the CRSP U.S. Stock Database for the period 1929-2022. \\}
\label{Table:GeneralPricesFunction_gamma_theta2}
\end{table}

\end{document}